\documentclass[prd,aps,preprintnumbers]{revtex4}

\usepackage{epsfig,subfigure}
\usepackage{amsmath}
\usepackage{epstopdf}

\def\P{{\mathcal P}}

\def\a0{\alpha_0}

\def\vep{\varepsilon}

\def \ep{\epsilon}

\def \nn {\nonumber}

\def\P {{\cal P}}

\def\bea {\begin{eqnarray}}
\def\eea {\end{eqnarray}}

\def\be {\begin{equation}}
\def\ee {\end{equation}}
\def\bi {\begin{itemize}}
\def\ei {\end{itemize}}

\begin{document}

\preprint{YITP-SB-17-15}

\renewcommand{\thefigure}{\arabic{figure}}

\title{Path description of coordinate-space amplitudes}

\author{Ozan Erdo\u{g}an}

\affiliation{Department of Physics, Carnegie Mellon University, Pittsburgh, PA 15213, USA}

\author{George Sterman}

\affiliation{C.N.\ Yang Institute for Theoretical Physics and Department of Physics and Astronomy\\
Stony Brook University, Stony Brook, NY 11794--3840, USA}

\date{\today}

\begin{abstract}
We develop a coordinate version of light-cone-ordered perturbation theory, for general time-ordered products of fields, by carrying out integrals over one light-cone coordinate for each interaction vertex.   The resulting expressions depend on the lengths of paths, measured in the same light-cone coordinate.   Each path is associated with a denominator equal to a ``light-cone deficit", analogous to the ``energy deficits" of momentum-space time- or light-cone-ordered perturbation theory.   In effect,  the role played by intermediate states in momentum space is played by paths between external fields in coordinate space.   We derive a class of identities satisfied by coordinate diagrams, from which their imaginary parts can be derived.   Using scalar QED as an example, we show how the eikonal approximation arises naturally when the external points in a Green function approach the light cone, and we give applications to products of Wilson lines.   Although much of our discussion is directed at massless fields in four dimensions, we extend the formalism to massive fields and dimensional regularization.

\end{abstract}

\maketitle

\section{Introduction}

Perturbative coordinate-space integrals arise naturally in the evaluation of the expectation values of Wilson lines and loops, which order fields in position space \cite{Polyakov:1980ca,Erdogan:2011yc}.   The analytic structure of these expectations has been discussed recently in Refs.\ \cite{Laenen:2014jga,Laenen:2015jia}, which trace the origin in coordinate space of their imaginary parts.   In general, of course, the singularities of Green functions are of interest, and  in \cite{Erdogan:2013bga,Erdogan:2014gha} we analyzed position-space singularities in Green functions with massless lines at configurations related to physical scattering processes, finding a close correspondence to momentum-space results in terms of Landau equations \cite{Landau:1959fi} and their corresponding physical interpretations \cite{Coleman:1965xm,Sterman:1978bi}.   

Here we continue these investigations by developing a coordinate version of light-cone-ordered perturbation theory (LCOPT), long familiar from its use in momentum space.  Our aim is to provide another tool to illustrate all-orders properties of the generic weak-coupling expansion of Green functions in coordinate space.   Historically, coordinate-space formulations of  operator products and Green functions led to the operator product expansion and  Wilsonian formulations of the renormalization group.    While most current interest in these methods is their nonperturbative applications, coordinate analysis at weak coupling also finds applications to theories beyond Minkowski space \cite{Maldacena:2015iua,Aharony:2016dwx}.

The use of light-cone coordinates for perturbative integrals in momentum space has a long history and many applications \cite{Weinberg:1966jm,Brodsky:1997de}.    One of these is the representation of diagrams in light-cone-ordered form after the integration over (for example) the minus components of all loop momenta \cite{Chang:1968bh,Kogut:1969xa}.   In this paper, we find an analogous construction for Green functions in coordinate space, with important similarities and interesting differences compared to the momentum-space formalism.   In particular, our results will enable us to interpret the relationship between coordinate-space propagator poles and imaginary parts found for the expectations of Wilson lines in Refs.\ \cite{Laenen:2014jga,Laenen:2015jia}.    We will find that in coordinate amplitudes, imaginary parts are associated with particular paths through diagrams, sequences of propagators that connect external points by a set of light-cone vectors.    This is dual to the picture of momentum-space imaginary parts, which are associated with on-shell states of particles.   Most of our discussion is specific to four dimensions with massless lines, but the 
generalizations to dimensional regularization and massive fields are sketched as well.

To set the stage, we begin Sec.\ \ref{sec:coord-lc} with a brief review of basic formulas for LCOPT in momentum space, and their relation to unitarity, as a preface for
the main analysis of the paper.     Turning to coordinate space, we propose general formulas for light-cone integrals over the positions of internal vertices, illustrated by an example that shows the role of paths.  
In Sec.\ \ref{sec:path-denominators} we show that a construction involving paths that always flow forward in light-cone time is applicable to arbitrary diagrams and confirm the generality of the results of Sec.\ \ref{sec:coord-lc}.   We go on to applications in Sec.\ \ref{sec:eikonal-Wilson}, deriving the eikonal approximation for the scalar QED vertex, and relating our formalism to Wilson lines.   In Sec.\ \ref{sec:Disc-and-imaginary-for-Wilsonline} we study discontinuities in coordinate-space integrals, and rederive and reinterpret the results of  Refs.\ \cite{Laenen:2014jga,Laenen:2015jia}.   Finally, a brief discussion of the extension to dimensional regularization and massive fields is given as Sec.\ \ref{sec:mass-dim}, followed by a summary.   In an appendix, we offer an alternate derivation of the coordinate path expressions of Sec.\ \ref{sec:path-denominators}.

\section{Coordinate Light-Cone Ordering and Paths}
\label{sec:coord-lc}

To set the context, we summarize the rules of light-front perturbation theory in momentum space \cite{Chang:1968bh,Kogut:1969xa} in a brief subsection, emphasizing the roles of states and their connection to unitarity.   In the subsequent discussion, we show how similar expressions can be derived from coordinate-space integrals, and how paths occur as the basic organizing principle.   Working from a specific example, we identify ordered paths, and describe the basic construction.    The demonstration that this construction applies to arbitrary diagrams and orders is the subject of the following section.

\subsection{Light-cone-ordered perturbation theory in momentum space}
\label{sec:review}

We imagine a momentum-space Green function,  $G\left (\{ l_d\},\{k_c\} \right )$ with $m$ incoming momenta $k_c$ and $n$ outgoing momenta $l_d$, so that $E\equiv m+n$ is the number of external lines.  For any covariant (Feynman) diagram corresponding to this Green function, we can integrate over the minus components of all loop momenta. The result is a perturbative expansion in terms of plus-momentum deficits, which has many fewer terms than the time-ordered form that results from energy integrals.   A diagram $G$ can be written as a sum of terms corresponding to orderings, $\P$ of its vertices, in which the plus momentum flows forward on every line. The diagram then describes a set of states, $s=1,\dots ,V-1$, where $V$ is the number of vertices in the diagram including external vertices, so that $V=E+N$, with $N$ the order. Each state corresponds to a cut of the ordered diagram between any two of its vertices.   The ordering of vertices determines the list of possible states.

Schematically, for a scalar diagram with lines of mass $m$, each such ordered diagram is of the form \cite{Chang:1968bh,Kogut:1969xa}
  \bea
-i\, G_\P\left (\{ l_d\},\{k_c\} \right )\ & = & \ g^N\; \int\left(\prod_{{\rm loops}\ i} \frac{dq^+_id^2q_{i\perp}}{(2\pi)^3}  \right) \prod_{{\rm lines}\ j} \frac{\theta(p_j^+)}{2p_j^+}\  
 \prod_{{\rm states}\ s=1}^{V-1}\   \frac{1}{ P_{\rm ext}^-{}^{(s)}\ - \sum_{p\in s} \frac{p_\perp^2+m^2}{2p^+}\ +\ i\ep }\, ,
 \label{eq:lcopt-mtm-basic}
\eea
where $P_{\rm ext}^-{}^{(s)}$ is the total external minus momentum flowing into the diagram before state $s$.  The overall factor of $-i$ on the left produces a contribution to the $T$ matrix after reduction, ensuring that below all thresholds the diagrams are real.   We suppress additional overall constants, including symmetry factors.

One of the convenient features of expressions like Eq.\ (\ref{eq:lcopt-mtm-basic}) is their close relation to unitarity.   In particular, the vanishing of any denominator corresponds to an on-shell state, often referred to as a cut of the (light-cone-ordered) diagram, which for state $J$ we denote by
\bea
G_\P^{(J)}\left (\{ l_d\},\{k_c\} \right )\ & \equiv &  \ g^N\; \int\left(\prod_{{\rm loops}\ i} \frac{dq^+_id^2q_{i\perp}}{(2\pi)^3}  \right) \prod_{{\rm lines}\ j} \frac{\theta(p_j^+)}{2p_j^+}\  
 \prod_{{\rm states}\ t=J+1}^{V-1}\   \frac{1}{ P_{\rm ext}^-{}^{(t)}\ - \sum_{p\in t} \frac{p_\perp^2+m^2}{2p^+}\ - \ i\ep }\
  \nn\\
  &\ & \hspace{5mm} \times\ 2\pi\; \delta\left( P_{\rm ext}^-{}^{(J)}\ - \sum_{p\in J} \frac{p_\perp^2+m^2}{2p^+}\  \right)
 \prod_{{\rm states}\ s=1}^{J-1}\   \frac{1}{ P_{\rm ext}^-{}^{(s)}\ - \sum_{p\in s} \frac{p_\perp^2+m^2}{2p^+}\ +\ i\ep }\, .
 \label{eq:momentum-cut}
 \eea
By repeated use of the identity $2i\pi\delta(x)=(x-i\epsilon)^{-1}-(x+i\epsilon)^{-1}$, we derive that the sum over all 
states $J$ of a given ordered diagram produces the imaginary part of the same ordered diagram, \cite{Sterman:1978bi}

\bea
2i\  {\rm Im} \ \left[ -iG_\P \left (\{ l_d\},\{k_c\} \right) \right]\ & = &\ \sum_{J=1}^{V-1}\ G_\P^{(J)}\left (\{ l_d\},\{k_c\} \right )\, .
\label{eq:momentum-unitarity}
\eea
This is a realization of unitarity on a point-by-point basis in the phase space of particles in the theory, which is helpful in a variety of physical situations.   For momentum space, this form applies to particles of any mass, and can easily be continued away from four dimensions.

\subsection{Light-cone-ordered perturbation theory in coordinate space}
\label{sec:pt-coord-subsec}

It is natural ask whether there is a useful ``dual" formalism for coordinate-space Green functions, which are a fundamental element in quantum field theory, and which also occur naturally in the evaluation of expectation values of Wilson lines \cite{Laenen:2014jga,Laenen:2015jia}.   We will see that a program of perturbative light-cone integrals can indeed be carried out in coordinate space, following a pattern similar to that in momentum space.   

For the massless case in four dimensions, we will show how to write an arbitrary diagram as a sum of products of simple denominators.  Each denominator is associated with a path between external vertices rather than with a multiparticle state found by cutting the diagram.    In this case, the denominators represent an ``excess" of distance between a particular path relating vertices and the path length for a particle moving on the light cone.   Unlike the case of momentum space, massless lines in four dimensions lead to qualitatively simpler results than we will find with massive lines and/or dimensional regularization, but we will see that the formalism can be extended both to massive and dimensionally-regulated theories. 
Most of our discussion will be for scalar theories, but we will comment on including spin, and give applications to gauge theories in later sections.

\begin{figure}

\centering
\subfigure[]{
\includegraphics[height=4cm]{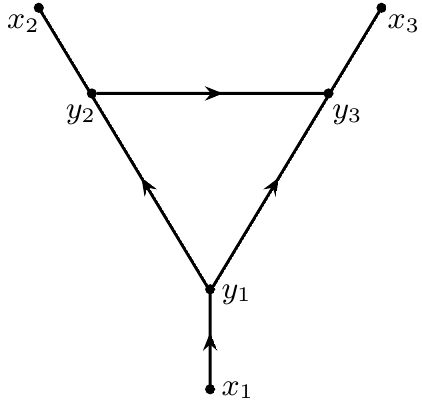}\label{fig:scalar-triA}}
\hspace{10mm}
\subfigure[]{
\includegraphics[height=4cm]{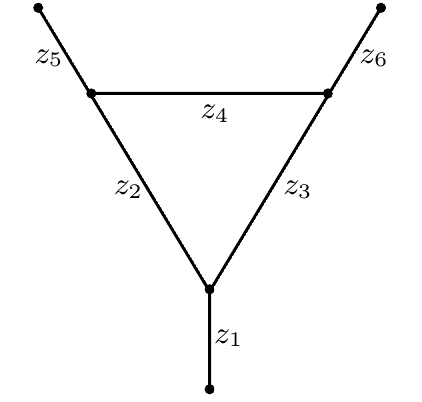}\label{fig:scalar-triB}}
\caption{Scalar triangle with external lines discussed in the text, showing \subref{fig:scalar-triA} labels for vertices and sense of vectors between vertices and \subref{fig:scalar-triB} labels for vectors between vertices.}  \label{fig:scalar-tri}

\end{figure}

We begin, then, with all massless scalar lines and in four dimensions.   
To illustrate our method, we will use the scalar triangle with three external lines, shown in Fig.\ \ref{fig:scalar-tri}.   In Fig.\ \ref{fig:scalar-triA} we label the vertices of the diagram and in Fig.~\ref{fig:scalar-triB} the lines, in a notation that we will use below.

For any diagram, the massless scalar propagator ($\Delta\equiv i\Delta_F$ here) is given by
\bea
\Delta(x^2)\ =\  \frac{1}{4\pi^2}\, \frac{1}{-x^2+i\ep}\, .
\label{eq:4d-massless-propagaor}
\eea
We analyze the general expression for an arbitrary connected scalar diagram with $E$ fixed external points in coordinate representation 
\be 
G_E(\{ x_a\}) \ = \  \frac{(-ig)^N}{(4\pi^2)^L}\; \int\prod_{\mathrm{vertices}\, i=1}^N d^4y_i \prod_{\mathrm{lines}\, j=1}^L \frac{1}{-z^2_j(y_i,x_a) + i\ep} \ ,
\label{eq:G-massless} 
\ee
where the $x_a$, $a=1, \dots , E$ denote external points and the $y_i$, $i=1, \dots , N$, internal vertices, and the vectors $z^\mu_j$ are defined by
\bea
z^\mu_j=(\eta_{ji}y_i+\eta'_{ja}x_a)^\mu\, .
\label{eq:z-etas}
\eea
Each $\eta_{ji}$ takes the value  $+1\, (-1)$ when $z_j$ is defined to end (begin) at vertex $i$, and is zero otherwise, and similarly for $\eta'_{ja}$ in terms of external vertices $x_a$.  Each $z_j^2$ then gives an invariant separation between the vertices connected by line $j$. Inserting the identity into the integrand for the minus component of each line difference, we write
\bea 
G_E\left (\{ x_a\} \right) & = &  \frac{(-ig)^N}{(4\pi^2)^L}\; \int\prod_{i\in N} d^4y_i \prod_{j\in L} \int dz^-_j\,\frac{1}{-2z^-_jz^+_j+z^2_{j\perp}+i\ep}\,
\delta \left (z^-_j -\eta_{ji}y^-_i - \eta'_{ja}x^-_a \right )  \nonumber \\
  & = &  \frac{(-ig)^N}{(4\pi^2)^L} \; \int\prod_{i\in N} d^4y_i \prod_{j\in L} 
  \int_{-\infty}^\infty \frac{dE^+_j}{2\pi} 
  \int_{-\infty}^\infty dz^-_j \,
  \frac{e^{iE^+_j \left (z^-_j -\eta_{ji}y^-_i - \eta'_{ja}x^-_a \right )} }{-2z^-_jz^+_j+z^2_{j\perp}+i\ep} \ , 
  \label{eq:delta-to-E}
\eea
where in the second equality we reexpress each delta function as an integral over a conjugate light-cone momentum, which we denote by $E^+_j$ and refer to as a {\it light-cone energy}.  Here and below, we will use the notation $N,E,V,L$ to denote the sets of internal vertices, of external vertices, of internal plus external vertices, and of lines of the diagram, and also the numbers of elements in each of these sets.   For the example in Fig.\ \ref{fig:scalar-tri}, $N=E=3$, $L=V=6$ and, for example, $\eta_{11}=1,\, \eta'_{11}=-1$, while $\eta_{43}=1,\, \eta_{42}=-1$, so that in particular, $z_4=y_3-y_2$, as indicated by the arrows in the figure.  

The $z_j^-$ contours in Eq.\ (\ref{eq:delta-to-E}) can be closed in the upper (lower) half plane for $E_j^+$ positive (negative), where they encounter poles only when $z_j^+$ has the same sign as $E_j^+$, a result familiar from light-cone-ordered perturbation theory in momentum space.   We can express this general result as
\bea 
G_E\left (\{ x_a\} \right ) 
& = & \  
 \frac{(-ig)^N}{(4\pi^2)^L}\; (-i)^L\int\prod_{i\in N} d^4y_i \prod_{j\in L} \int_{-\infty}^\infty dE^+_j\,e^{-iE^+_j(\eta_{ji}y^-_i + \eta'_{ja}x^-_a)}\,
\theta \left (z^+_jE^+_j \right )\frac{e^{iE^+_j\frac{z^2_{j\perp}+i\ep}{2z^+_j}}}{2|z^+_j|}  \nonumber \\
 &\ & \hspace{-20mm} =\
 (2\pi)^{N}\,  \frac{(-ig)^N}{(4\pi^2)^L} (-i)^L\int\left(\prod_{i\in N} d^3y_i\right) \prod_{j\in L} \int_{-\infty}^\infty dE^+_j\, \frac{e^{-iE^+_j \left (\eta'_{ja}x^-_a-\frac{z^2_{j\perp}+i\ep}{2z^+_j} \right ) }}{2|z^+_j|}\,\theta \left (z^+_jE^+_j \right ) \prod_{i\in N}\delta \left (E^+_j\eta_{ji} \right ) \, ,
 \label{eq:E-integrals-1}
 \eea
where in the second equality, we have integrated over the minus components of all internal vertices to obtain a momentum-conservation delta function for each internal vertex.   
Many of the basic features of light-cone ordering are already exhibited in Eq.\ (\ref{eq:E-integrals-1}).

The step functions in Eq.\ (\ref{eq:E-integrals-1}) show that each light-cone energy $E^+_j$ flows in the same direction as the corresponding plus component of line vector $z_j^\mu$.  This is just as in light-cone-ordered perturbation theory in momentum space, although here, of course, the positions of vertices are the integration variables.      At fixed $y^+_j$, the $E^+_j$ flow only ``forward" in any diagram once we have ordered its vertices from smallest to largest values of $y_i^+$, and they are conserved at each vertex.   Consequently, there are no contributions whenever any $y_i^+$ is either earlier or later than all vertices to which it is connected (internal or external).   That is, as in the momentum light-cone formalism, no sets of lines ``emerge from or disappear into the vacuum".

As a result, every internal vertex must have momentum flowing in from smaller $y^+$, and flowing out toward larger $y^+$.   All light-cone energies then flow in through a set of ``initial" external vertices $x_a$, and out through another set of ``final" external vertices.   Initial vertices $x_c$ are those that connect to internal vertices with $y_i^+>x_c^+$, and final vertices $x_d$ are those which connect to internal vertices with $y_i^+<x_d^+$.   Clearly, there must be at least one initial and at least one final vertex.   Every diagram is a sum over nonempty and distinguishable choices of sets for these vertices, determined by the $E$ step functions in Eq.\ (\ref{eq:E-integrals-1}) for lines $z_j$ that are connected to external vertices.   We may represent this sum as 
\bea
G_E\left (\{ x_a\} \right ) \ =\ \sum_{\{x_d\}_{\rm out},\{x_c\}_{\rm in} \ne \phi}\; 
G_{(n,m)}\left (\{x_d\}_{\rm out},\{x_c\}_{\rm in}\right )\, ,
\eea
with $n+m=E$.   Each term $G_{(n,m)}$ is of the form
\bea
G_{(n,m)} \left (\{x_d\}_{\rm out},\{x_c\}_{\rm in}\right )\ &=&\  (2\pi)^N\, \frac{(-ig)^N}{(4\pi^2)^L}\, (-i)^L\int\left(\prod_{i\in N} d^3y_i\right) \prod_{j\in L} \int dE^+_j\, \prod_{d=1}^n\theta(E^+_d)\ \prod_{c=1}^m\theta(E^+_c)
\nn\\
&\ & \hspace{20mm} \times
\frac{e^{-iE^+_j \left (\eta'_{ja}x^-_a-\frac{z^2_{j\perp}+i\ep}{2z^+_j} \right ) }}{2|z^+_j|}\,\theta \left (z^+_jE^+_j \right ) \prod_{i\in N}\delta \left (\eta_{ji}E^+_j \right ) \ ,
\label{eq:E-int-in-fin}
\eea
where we have adopted the notation that $E^+_c$ is the momentum flowing {\it  out of initial vertex} $x_c$ and $E^+_d$ is the momentum flowing {\it  into final vertex} $x_d$.   

For Fig.\ \ref{fig:scalar-tri}, there are six nonvanishing choices for the sets of  incoming and outgoing vertices.    We will consider one of these, with one incoming and two outgoing vertices, corresponding to the term
\bea
G_{(2,1)}\left (\{x_2,x_3\}_{\rm out},\{x_1\}_{\rm in}\right )\ &=&\  (2\pi)^3\, \frac{(-ig)^3}{(4\pi^2)^6}\, (-i)^6\int\left(\prod_{i=1}^3 d^3y_i\right) \prod_{j=1}^6 \int dE^+_j\, \prod_{d=2,3}\theta(E^+_d)\ \theta(E^+_1)
\nn\\
&\ & \hspace{20mm} \times
\frac{e^{-iE^+_j \left (\eta'_{ja}x^-_a-\frac{z^2_{j\perp}+i\ep}{2z^+_j} \right ) }}{2|z^+_j|}\,\theta \left (z^+_jE^+_j \right ) \prod_{i=1}^3\delta \left (\eta_{ji}E^+_j \right ) \, .
\label{eq:E-int-in-fin-example}
\eea

\begin{figure}

\centering
\includegraphics[height=4cm]{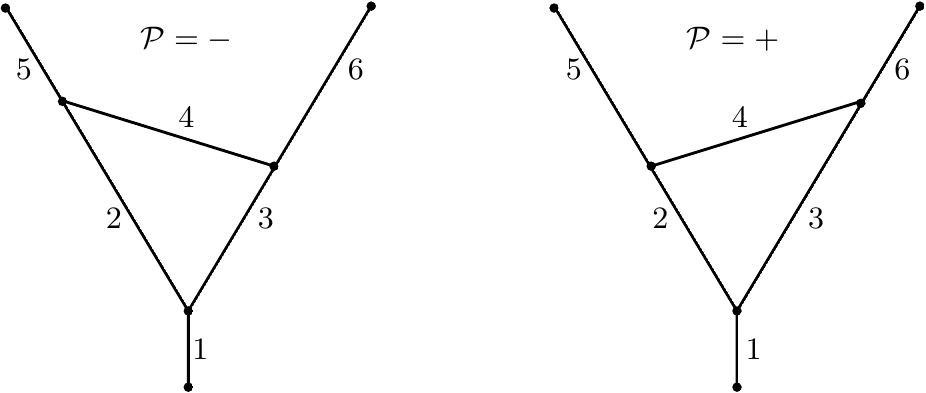}
\caption{Ordered diagrams discussed in text.}  
\label{fig:ordered-scalar-tri}

\end{figure}

As the next step in the general case, we insert unity in the form of all possible signs for each of the $z_j^+$, $j\in \{L\}$,
\bea
1\ &=&\ \prod_{j\in \{L\}}\ \left[ \theta\left(z_j^+\right)\ + \theta\left(-z_j^+\right) \right]
\nn\\
&\equiv&\ \sum_\P\ \prod_{j\in \P^+} \theta\left(z_j^+\right)\  \prod_{j\in \P^-} \theta\left(-z_j^+\right)
\nn\\
&\equiv& \ \sum_\P\ \Theta_\P\left (\{z_j\}\right)\, ,
\label{eq:partition-unity}
\eea
where $\P\equiv \P^+ \cup \P^-$ is the set of all $z_j$'s. Note that  in each
$G_{(n,m)}\left (\{x_d\}_{\rm out},\{x_c\}_{\rm in}\right )$ only one step function can contribute for the external lines, whose directions have been fixed by the choice of incoming and outgoing external vertices.     Then every term in Eq.\ (\ref{eq:E-int-in-fin}) itself becomes a sum,
\bea
G_{(n,m)}\left (\{x_d\}_{\rm out},\{x_c\}_{\rm in}\right ) \ =\ \sum_\P G_{(n,m)\P}\left (\{x_d\}_{\rm out},\{x_c\}_{\rm in}\right )\, ,
\eea
where
\bea
G_{(n,m)\P}\left (\{x_d\}_{\rm out},\{x_c\}_{\rm in}\right )\ &=&\  (2\pi)^N\, \frac{(-ig)^N}{(4\pi^2)^L}(-i)^L\int\left(\prod_{i\in N} d^3y_i\right) \ \Theta_\P\left (\{z_j\}\right)\
\prod_{j\in L} \int dE^+_j\, \prod_d\theta(E^+_d)\ \prod_c\theta(E^+_c)
\nn\\
&\ & \hspace{20mm} \times\
\frac{e^{-iE^+_j \left (\eta'_{ja}x^-_a-\frac{z^2_{j\perp}+i\ep}{2z^+_j} \right ) }}{2|z^+_j|}\,\theta \left (z^+_jE^+_j \right ) \prod_{i\in N}\delta \left (\eta_{ji}E^+_j \right ) \, .
\label{eq:E-int-F_P}
\eea
For each choice of $\P$ we may now change variables to {\it redefine} the $z_j$'s so that all of them are positive.   This requires us to also change the sign of those $E_j^+$'s whose $z_j^+$'s were negative.    We keep the same notation for all the integrals, but observe that the effect of these changes can be absorbed entirely into the incidence matrix elements $\eta_{ji}$ and $\eta'_{ja}$, which simply change sign whenever $E_j^+$ changes sign.  Note that $E^+_a$ does not change sign when $\eta'_{j_aa}\neq 0$, but the original integration variable for that line, $E^+_{j_a}$ may change sign. The choice of partition $\P$ then, is encoded entirely in the new incidence matrices, which we denote by $\eta_{ji}^{(\P)}$ and $\eta_{ji}'{}^{(\P)}$.    We may thus write
\bea
G_{(n,m)\P}\left (\{x_d\}_{\rm out},\{x_c\}_{\rm in}\right )\ &=&\  (2\pi)^N\, \frac{(-ig)^N}{(4\pi^2)^L}\,  (-i)^L\int\left(\prod_{i\in N} d^3y_i\right) \ \prod_{{\rm all}\ j} \theta(z_j^+)\
\prod_{j\in L} \int dE^+_j\, \
\nn\\
&\ & \hspace{20mm} \times
\frac{e^{-iE^+_j \left (\eta'_{ja}{}^{(\P)}x^-_a-\frac{z^2_{j\perp}+i\ep}{2z^+_j} \right ) }}{2z^+_j}\,\theta \left (E^+_j \right ) \prod_{i\in N}\delta \left (\eta^{(\P)}_{ji}E^+_j \right ) \ .
\label{eq:E-int-F_P-positive}
\eea
For our example, Eq.\ (\ref{eq:E-int-in-fin-example}), with the choice of $z_i^\mu$ as in Fig.\ \ref{fig:scalar-tri}, we consider two terms from the sum in Eq.\ (\ref{eq:partition-unity}) that we label $\P=4^+$ and $\P=4^-$, given  by
$\theta(z_2)\theta(z_3)\theta(z_4)$ and $\theta(z_2)\theta(z_3)\theta(-z_4)$.   In fact, these turn out to give the complete answer in this case.  
After making all the $z_j$'s positive, we have for $\P=4^+$ and $\P=4^-$ forms that differ only in the signs of energy $E_4^+$ in the light-cone energy conserving delta functions,
\bea
G_{(2,1)4^\pm}\left (\{x_2,x_3\}_{\rm out},\{x_1\}_{\rm in}\right )\ &=&\  (2\pi)^3\, \frac{(-ig)^3}{(4\pi^2)^6}\, (-i)^6\int\left(\prod_{i=1}^3 d^3y_i\right) 
\prod_{j=1}^6  \int \frac{dE^+_j} {2z^+_j}\, \theta(z_j^+)\, \theta(E^+_j)\  
\nn\\
&\ & \hspace{3mm} \times 
\ \delta \left (E^+_2-E^+_5 \mp E^+_4 \right )\, \delta \left (E^+_3-E^+_6 \pm E^+_4 \right )\, \delta \left (E^+_1-E^+_3-E^+_2 \right )
\nn \\
&\ & \hspace{3mm}
\times\ e^{-i \left ( x_2^-\, E^+_5\ +\ x_3^-\, E^+_6\ -\ x_1^-\, E^+_1 \right)}\ e^{ i\left( \sum_{j=1}^6 E_j^+\, \frac{z^2_{j\perp}+i\ep}{2z^+_j} \right ) } 
 \, .
\label{eq:E-int-in-fin-example-4plus}
\eea
These results correspond to the two orderings of vertices shown in Fig.\ \ref{fig:ordered-scalar-tri}.   In these expressions, $z_4^+=\pm (y_3^+-y_2^+)$, with the minus referring to $G_{(2,1)4^-}$.

Our aim now is to perform the integrals over light-cone energies in Eq.\ (\ref{eq:E-int-F_P-positive}), with a given choice of initial and final vertices, to derive an expression in terms of integrals over the coordinates $\{y_i^+,y_{i\perp}\}$ only.   For an arbitrary diagram, this will require several steps, but we can get a good idea of the procedure by evaluating the special cases of Eq.\ 
(\ref{eq:E-int-in-fin-example-4plus}), first for $\P=4^+$,
\bea
G_{(2,1)4^+}\left (\{x_2,x_3\}_{\rm out},\{x_1\}_{\rm in}\right )\ &=&\  (2\pi)^3\, \frac{(-ig)^3}{(4\pi^2)^6}\, (-i)^6\int\left(\prod_{i=1}^3 d^3y_i  \prod_{j=1}^6 \frac{\theta(z_j^+)}{2z^+_j}\right)  
\nn\\
&\ & \hspace{3mm} \times  \int dE^+_6\,  \theta(E^+_6)\    
e^{-i E_6^+\, \left (   \left [x_3^--x_1^- \right ]\ -\  \left[ \frac{z^2_{1,\perp}}{2z^+_1} +\frac{z^2_{3,\perp}}{2z^+_3}+\frac{z^2_{6,\perp}}{2z^+_6}  \right] -i\ep \right ) }
\nn\\
&\ & \hspace{3mm} \times  \int dE^+_5\,  \theta(E^+_5)\  
e^{-i E_5^+\, \left (   \left [x_2^--x_1^- \right ]\ -\  \left[ \frac{z^2_{1,\perp}}{2z^+_1}+\frac{z^2_{2,\perp}}{2z^+_2} +  \frac{z^2_{5,\perp}}{2z^+_5} \right] -i\ep \right ) }
\nn\\
&\ & \hspace{3mm} \times
\int dE^+_4\, \theta( E^+_4)
\ \theta \left ( E^+_6-E^+_4 \right )\  e^{- i E_4^+ \left[ \frac{z^2_{3,\perp}}{2z^+_3}-\frac{z^2_{2,\perp}}{2z^+_2}-\frac{z^2_{4,\perp}}{2z^+_4}  \right] }  \, .
\label{eq:E-int-in-fin-example-5plus}
\eea
Secondly, for $\P=4^-$ we get
\bea
G_{(2,1)4^-}\left (\{x_2,x_3\}_{\rm out},\{x_1\}_{\rm in}\right )\ &=&\  (2\pi)^3\, \frac{(-ig)^3}{(4\pi^2)^6}\, (-i)^6\int\left(\prod_{i=1}^3 d^3y_i  \prod_{j=1}^6 \frac{\theta(z_j^+)}{2z^+_j}\right)  
\nn\\
&\ & \hspace{3mm} \times  \int dE^+_6\,  \theta(E^+_6)\ 
e^{-i E_6^+\, \left (   \left [x_3^--x_1^- \right ]\ -\  \left[  \frac{z^2_{1\perp}}{2z^+_1}+\frac{z^2_{3\perp}}{2z^+_3} +\frac{z^2_{6\perp}}{2z^+_6} \right] -i\ep \right ) }
\nn\\
&\ & \hspace{3mm} \times  \int dE^+_5\,  \theta(E^+_5)\   
e^{-i E_5^+\, \left (   \left [x_2^--x_1^- \right ]\ -\  \left[ \frac{z^2_{1\perp}}{2z^+_1}+\frac{z^2_{2\perp}}{2z^+_2} +  \frac{z^2_{5\perp}}{2z^+_5} \right]  -i\ep \right ) }
\nn\\
&\ & \hspace{3mm} \times
\int dE^+_4\, \theta( E^+_4) 
\ \theta \left ( E^+_5-E^+_4 \right )\  e^{- i E_4^+ \left[ \frac{z^2_{2\perp}}{2z^+_2}-\frac{z^2_{3\perp}}{2z^+_3}-\frac{z^2_{4\perp}}{2z^+_4}  \right] }  \, .
\label{eq:E-int-in-fin-example-5minus}
\eea
We now integrate over the remaining light-cone energies and obtain an expression in terms of denominators corresponding to each path from the initial external vertex to a final external vertex. For $\P=4^+$, the result is given by
\bea
G_{(2,1)4^+}\left (\{x_2,x_3\}_{\rm out},\{x_1\}_{\rm in}\right ) &=&  (2\pi)^3\, \frac{g^3}{(4\pi^2)^6}\, \int\left(\prod_{i=1}^3 d^3y_i  \prod_{j=1}^6 \frac{ \theta(z_j^+)}{2z^+_j}\right) \nn \\
& \ & 
\hspace{-20mm}\times\frac{1}{ x_2^--x_1^- -  D_{521}
 -i\ep} \,
\frac{1}{ x_3^--x_1^-  - D_{631}  
 -i\ep } \,\frac{1}{ x_3^--x_1^-  - D_{6421}
  -i\ep } \ , 
\label{eq:path-form-example-5plus}
\eea
where each denominator corresponds to a different path going forward from the initial vertex to either of the final vertices, and where we have defined the cumulative ``light-cone distance'' for each line traversed on a path,
\be D_{i_n,\dots , i_1} = \sum^n_{j=1}     \frac{z^2_{i_j\perp}}{2z^+_{i_j}} \ . \label{eq:Dlines}\ee
Similarly, for $\P=4^-$, we find
\bea
G_{(2,1)4^-}\left (\{x_2,x_3\}_{\rm out},\{x_1\}_{\rm in}\right ) &=&  (2\pi)^3\, \frac{g^3}{(4\pi^2)^6}\, \int\left(\prod_{i=1}^3 d^3y_i  \prod_{j=1}^6 \frac{ \theta(z_j^+)}{2z^+_j}\right)  
\nn\\
&\ & \hspace{-20mm} \times\frac{1}{ x_3^--x_1^-  - D_{631}
 -i\ep } \, 
\frac{1}{ x_2^--x_1^- -  D_{521}
-i\ep } \, \frac{1}{ x_2^--x_1^-  -  D_{5431}
-i\ep} \ .
\label{eq:path-form-example-5minus}
\eea
In the next section, we will construct an algorithm to perform the integrals over light-cone-energy variables, generalizing these results to generic Green functions.  We will find that any diagram can be expressed as a sum of terms involving the same  path denominators found in our examples,
  \be
G^{(\pi)}_{(n,m)\P}\left (\{x_d\}_{\rm out},\{x_c\}_{\rm in}\right ) \ =\
  (2\pi )^{N-2L}\, (-g)^N  \int\left(\prod_{i\in N} d^3y_i\right) \prod_{j \in L} \frac{\theta(z_j^+)}{2z_j^+} 
 \prod_{\{P^{(\pi)}_{(ba)}\} \in \pi} \frac{-1}{x^-_b\ -\ x^-_a\ -\ D^{(\pi)}_{(ba)} -i\ep }\, ,
 \label{eq:1st-result-massless}
\ee 
where each $P^{(\pi)}_{(ba)} \in \pi$ denotes a path extending from some initial vertex $x_a\in \{x_c\}_{\rm in}$ to some final vertex $x_b\in \{x_d\}_{\rm out}$, and the union of all these paths, $\cup P^{(\pi)}_{(ba)} \equiv \pi$ covers the diagram.   As we shall see, the choice of paths is not unique, but can be made well defined.
The terms $D^{(\pi)}_{(ba)}$ are the ``light-cone distances" associated with a path from $x_a$ to $x_b$, defined by the plus and transverse components of the intermediate vertices,
\bea
D^{(\pi)}_{(ba)}\ =\ \sum_{{\rm vertices}\ i\in P^{(\pi)}_{(ba)}} \frac{z_{i,i-1,\perp}^2}{2z_{i,i-1}^+}\, ,
\label{eq:1st-deficit-def}
\eea
where $z^\mu_{i,i-1}$ is the distance between the $i-1$st and $i$th vertices in the path, ordered by plus component, with $x_a$ the earliest vertex and $x_b$ and the latest vertex.  For the examples above, each ordering has one such set consisting of three paths that extend from $x_1$ to $x_2$ or $x_3$, and  $D^{(\pi)}_{(ba)}$ are given by $ D_{i_n,\dots , i_1}$ in Eq.~(\ref{eq:Dlines}).  
In the following section, we will derive the generalization of this result in terms of paths and distances at arbitrary order.   

\section{Path Denominators to All Orders}
\label{sec:path-denominators}

\subsection{Energy integrals in the general case}

As a first step in the general case, we use the light-cone-energy delta functions in Eq.\ (\ref{eq:E-int-F_P-positive}) to obtain an integral over the remaining $L-N$  light-cone energies.  The restrictions on the light-cone energies in (\ref{eq:E-int-F_P-positive}) are then that they combine to flow forward on every line, they are conserved at vertices, and that the
 total momentum flowing through initial and final vertices are equal.   We can always choose these independent momenta to flow through the diagram from the initial external points to the final points.   They can be thought of as a set of independent loops for the vacuum bubble diagram found by combing all external vertices into a single vertex.   When $G$ is a tree diagram, in particular, $L-N=E-1=m+n-1$, where again $L$ is the total number of lines and $m$ ($n$) the number of initial (final) vertices. Each internal loop of $G$ simply adds one additional integral.    

To begin, we choose the integration variables as the light-cone energies carried by the lines connected to the external vertices, denoted $k_i$, $i=1, \dots ,  m$ for the lines connected to initial vertices and 
$l_j$, $j=m+1,  \dots , m+ n$ for the lines connected to final vertices.  Applied to these external lines, the presence of a step function for every line in Eq.\ (\ref{eq:E-int-F_P-positive}) coupled with energy conservation implies that the light-cone energy $E_j^+$ carried by any line $j$ always flows forward from smaller values of $y^+$ to larger values.   Then, every internal vertex, at $y_i^+$, say, must connect to at least one vertex $y_j^+>y_i^+$, and at least one vertex $y_{j'}$ with $y_{j'}^+< y_i^+$.       For simplicity, we assume that every external vertex is connected to an internal vertex by a single line, that is, that the diagram represents a perturbative contribution to a time-ordered product of single fields.  Each external vertex is then connected to a single internal vertex by a single line.    Note that the $S$-matrix elements that describe $m\rightarrow n$ scattering are found from the Fourier transforms of $G_{(n,m)}$.

For tree diagrams, the light-cone energies of external lines determine the light-cone energies of all internal lines.  Contributions to products of composite operators can be found by identifying two or more external vertices.   Diagrams with loops require more light-cone energies.   

In summary, from the step functions in Eq.\  (\ref{eq:E-int-F_P-positive}), positive energy variables $l^+_j$ always flow out of the diagram into final vertices, $X_j$ while positive energies $k^+_i$  flow out of the initial vertices $x_i$ into the diagram.   For any such relative orderings between internal and external vertices, the full space of external light-cone energies is then given by
\bea
k^+_i\ &\ge & 0\, , i\ =\ 1, \dots , m\, , \nn\\
l_j^+\ &\ge & 0\, ,  j= m+1, \dots ,  m+n\, ,
\label{eq:k-l-notation}
\eea
subject to overall momentum conservation, requiring (only) that
\bea
\sum_{i=1}^{m} k_i\ \ =\ \sum_{j=m+1}^{m+n} l_j \, .
\eea
We denote the remaining (loop) momenta as $p_s$.  
In these terms, we may rewrite the general integral Eq.\ (\ref{eq:E-int-F_P-positive}), assuming $K$ loops, as
  \bea 
  G_{(n,m)\P}\left (\{x_d\}_{\rm out},\{x_c\}_{\rm in}\right ) & = & (2\pi)^N\, \frac{(-ig)^N}{(4\pi^2)^L}\, (-i)^L\int\left(\prod_{i\in N} d^3y_i\right) 
  \prod_{j=m+1}^{m+n} \int_0^\infty dl_j\, \prod_{i=1}^m \int_0^\infty dk_i\, \prod_{s=1}^K \int_{-\infty}^\infty dp_s\
  \nn\\
  &\ & \hspace{10mm} \times\ \delta \left ( \sum_{j=m+1}^{m+n} l_j\ \ -\ \sum_{i=1}^{m} k_i  \right )
  \nn\\
&\ & \hspace{10mm} \times
  \prod_{w=1}^L
  \frac{\theta(z_w^+)\theta \left ( q^+_c\sigma^{(\P)}_{cw} \right ) }{2z^+_w}\, 
  \exp\left(-iq^+_c\sigma^{(\P)}_{cw} \left (\eta'_{wa}{}^{(\P)} x^-_a-\frac{z^2_{w\perp}+i\ep}{2z^+_w}\right ) \right) \ ,
  \label{eq:E-integrals-loops}
  \eea
  with $\sigma_{lj}^{(\P)}$ the incidence matrix for light-cone energies $l_j$, $k_i$ and $p_s$, collectively denoted $q^+_c$ here, along lines $w$ for this ordered diagram, $G_{(n,m)\P}$.   Overall momentum conservation is taken into account through the explicit delta function.  
  
  Note that the loop momenta in Eq.\ (\ref{eq:E-integrals-loops}) are not yet constrained to be positive.    In the following section, we use the step functions of this expression to derive a set of  light-cone variables whose integration regions are all positive and independent.  
  The price will be a sum over terms.   In each of these terms, however,
  we will be able to perform the integrals and derive a closed form with products of denominators that are determined by paths from initial to final vertices, as in the examples of Sec.\ \ref{sec:pt-coord-subsec}.

\subsection{Partial ordering of vertices}

The structure of the diagram after minus component integrals, Eq.\ (\ref{eq:E-integrals-loops}) is clarified by a discussion of paths between initial and final vertices.   This is based on a natural partial ordering of internal and external vertices based on the plus components of their positions.  The relevance of partially-ordered sets (``posets") to diagrammatic calculations for Wilson lines particularly has been emphasized in Refs.\ \cite{Dukes:2013wa,Dukes:2013gea}.  

Suppose that line $z_{fe}=y_f-y_e$ connects vertices $y_f$ and $y_e$.   We will say that $y_f$ is a  ``descendant" of $y_e$, denoted  $y_f>y_e$, if $z^+_{fe}>0$.    Equivalently, if $y_f>y_e$, vertex $e$ is a ``precursor" of vertex $f$.  The same terminology can be extended to lines.  More generally, we say that external or internal vertex $y_h$ is a descendant of another vertex $y_g$ if it is connected by a path, all of the intermediate vertices of which are descendants of $y_g$.   From the product of step functions in the integral (\ref{eq:E-integrals-loops}), we have a set of easy and useful properties, listed here.

\begin{enumerate}

\item Light-cone energies flow only from precursor vertices or lines to their descendants.

\item Every internal vertex is descended from one or more other (initial or internal) vertices and has at least one descendant.

\item Every final vertex has no descendants.

\item Every final vertex is a descendant of at least one initial vertex.

\item Every initial vertex is the precursor of at least one final vertex.

\item The light-cone energy carried by any initial line, $k_i^+$ is less than or equal to the sum of all momenta carried by
its final-state descendants, $k_i^+\le \sum_{X_j>x_i}\ l_j^+$.

\end{enumerate}
 
We will identify sets of new, positive semi-definite integration variables, each of which follows a path that  starts at an initial vertex and ends at one of its descendant final vertices.    These paths will include every line of the diagram, and we will call each such set a ``positive covering" of the diagram, denoted $\pi_I$.  Each positive covering corresponds to a disjoint subspace of the integrations in Eq.\ (\ref{eq:E-integrals-loops}).    Taken together, we will term the group of all positive coverings a ``covering class", denoted ${\cal C}=\cup\, \pi_I$ for the diagram.  For a given ordered diagram $G_\P$, the union of the disjoint momentum ranges of the elements $\pi_I$ of the covering class ${\cal C}$ is the full space in (\ref{eq:E-integrals-loops}),
\bea
G_\P\left (\{x_d\}_{\rm out},\{x_c\}_{\rm in}\right )\ &=&\   \sum_{\pi_I \in {\cal C}} G_\P^{(\pi_I)}\left (\{x_d\}_{\rm out},\{x_c\}_{\rm in}\right )\, .
\eea   
As already observed, we aren't yet able to do the energy integrals as expressed in  the form of (\ref{eq:E-integrals-loops}) explicitly for the general case, because the step functions for each line determine the lower limits of each of the integrals in a manner that depends on the details of each diagram and the choice of loops.   We will show below, however, that we can write this expression as a sum of terms in which $m+n-1+K$ integrals are freely integrated from zero to infinity, resulting in a set of denominators that correspond to ``distance deficits", the coordinate analog of the energy deficits of momentum-space time-ordered perturbation theory.  We will provide a construction that produces what we referred to above as a sum over sets of paths, $\pi_I$ that cover the diagram.  Like the choice of loop momenta in momentum space, this construction involves some freedom, which we will describe.   For this purpose, we will use the ordering properties of the vertices listed above.

\subsection{Choice of light-cone variables}

We begin with tree diagrams ($K=0$ in Eq.\ (\ref{eq:E-integrals-loops})). Our aim is to change variables from the $m$ $k_i$ and the $n$ $l_j$ to a new set of $n+m-1$ energies $q_{fi}$, each of which is associated with a path (from initial vertex $i$ to final vertex $f$).  As above, the set of paths determined by each change of variables will be denoted $\pi_I$, and their union referred to as ${\cal C}$.   These new energies are linearly related to the original set by
\bea
k_c\ &=&\ \sum_b \lambda^{(\pi_I)}_{bc} q_{bc}\, , \nn\\
l_d\ &=&\ \sum_a \mu^{(\pi_I)}_{da} q_{da}\, ,
\label{eq:reexpress-momenta}
\eea
where the coefficients $\lambda_{bc}^{(\pi_I)}$ and $\mu_{da}^{(\pi_I)}$ equal either $1$ or $0$.   When $\lambda_{bc}^{(\pi_I)}=1$, light-cone energy 
$q_{bc}$ flows into the diagram through initial vertex $x_c$ and out through final vertex $x_b$, and when $\mu_{da}^{(\pi_I)}=1$, 
$q_{da}$ flows into the diagram through initial vertex $a$ and out  through final vertex $x_d$.   Every $q_{fi}$ will flow in through exactly one vertex and out through exactly one vertex, flowing forward in plus coordinate along 
a set of lines to a unique final vertex, so that for fixed initial index $a$ and final index $d$,
\bea
\lambda_{da}^{(\pi_I)}\ =\ \mu_{da}^{(\pi_I)}\, .
\label{eq:lambda-mu-equality}
\eea
 In a tree diagram, the path between any pair of vertices is determined uniquely by the ordering of vertices, $\P$.   In a loop diagram, there may be more than one path between a given initial and final vertex, and it may be necessary to introduce more than one $q_{fi}$ between a given pair of vertices.   In fact, this is already the case for the one-loop example described in Sec.\ \ref{sec:pt-coord-subsec}.

The construction of the new variables $q_{fi}$ is a rather general problem, which, however, is simple to state.   We imagine two sets of positive variables, whose sums are equal:  $w_1+w_2 + \dots + w_m= z_1+z_2 +\dots + z_n$, which we wish to integrate from zero to infinity.   We would like to replace this constrained integral over these $n+m$ variables with a sum of integrals over $n+m-1$ variables that are integrated freely from zero to infinity.    It turns out that this requires us to sum over a number of terms, each with different changes of variables to cover the full space, and we will give below a prescription for constructing acceptable sets of new variables.    For us, each new variable will correspond to a path.

The motivation for changing variables in Eq.\ (\ref{eq:E-integrals-loops}) to the $q_{fi}$ according to (\ref{eq:reexpress-momenta})   is that every element in the sum over covering sets $\pi_I$ provides a term in which each energy integral  $q_{fi}$ is independently integrated from zero to infinity, giving, as in the examples above and Eq.\ (\ref{eq:1st-result-massless}), a denominator in the form of a light-cone deficit.  

\subsection{Example}

\begin{figure}

\centering
\subfigure[]{\includegraphics[height=4cm]{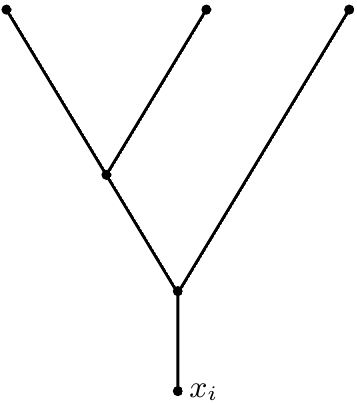}\label{fig:tree-figA}}
 \hspace{20mm}  
\subfigure[]{\includegraphics[height=4cm]{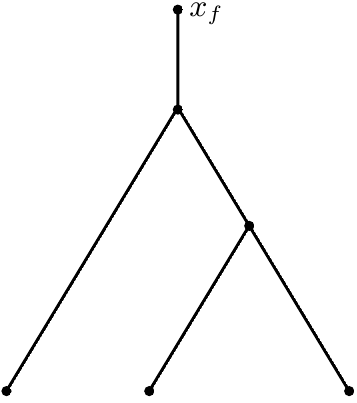}\label{fig:tree-figB}}

\caption{(a) An ordered tree diagram with a single initial vertex. (b) An ordered tree diagram with a single final vertex.  \label{fig:tree-fig}}

\end{figure}

For some orderings, the choice of light-cone energies, $q_{fi}$ is quite straightforward.   Simple (but completely representative) examples that illustrate such cases are shown in Fig.\ \ref{fig:tree-fig}.   Figure \ref{fig:tree-figA}, for example, shows a tree subdiagram for which all  final vertices are descendants of only a single initial vertex, $x_i$.  In this case, we can choose one integration variable as $q_{di}=k_d$, 
the light-cone energy on the line connected to each final vertex, $x_d$, which flows into the ordered diagram from $x_i$ and to $x_d$.   In the notation of of Eq.\ (\ref{eq:reexpress-momenta}), this corresponds to $\lambda^{(\pi)}_{di}=\mu^{(\pi)}_{di}=1$.   There is only a single path between vertex $x_i$ and each $x_d$, so the choice of momentum flow is unique, and each of the energies $q_{di}$ is independent, taking on any value between zero and infinity.    
Here, there is only a single covering set, $\pi$.

The choice of energy flows in a diagram like Fig.\ \ref{fig:tree-figB}, where a single final vertex ($x_f$) is the only descendant of a set of one or more initial vertices, is essentially identical. Again, there is only a single covering set, $\pi$.    There is one light-cone momentum, $q_{fb}$ for each initial vertex $x_b$, which flows along a determined path to the final vertex $x_f$.  Again, these momenta are freely integrated from zero to infinity.   The situation is more complex, however, whenever a set of final vertices has more than one predecessor, so that several final vertices are descendants of a set of initial vertices.

\begin{figure}

\centering
\subfigure[]{
\includegraphics[height=4cm]{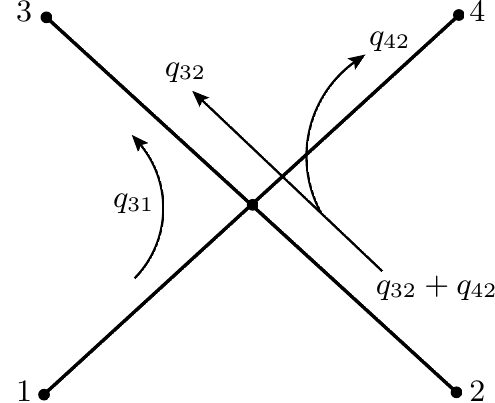}\label{fig:4pexA}}
\hspace{20mm}
\subfigure[]{
\includegraphics[height=4cm]{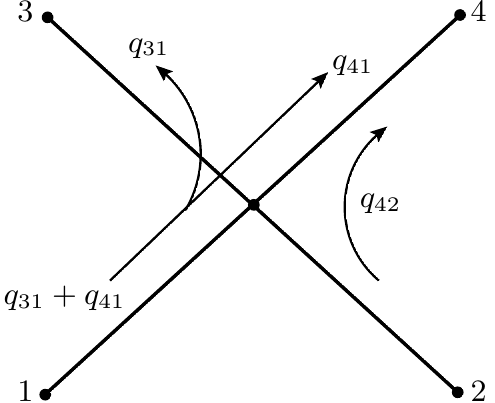}\label{fig:4pexB}}

\caption{Light-cone-energy flows, given in Eq.\ (\ref{eq:reexpress-momenta-4point}) for the lowest-order four-point vertex.  \label{fig:4-point-example}}

\end{figure}

The simplest example that illustrates the ambiguity associated with the ordering among final and initial vertices within a single diagram is the four-point function with a single four-point internal vertex, $y$, as in Fig.\ \ref{fig:4-point-example}.   This example is by itself completely trivial as an expression, but still serves to introduce the general case.  We start with four integration variables, which we integrate over all positive values subject to the constraint $k_1+k_2=l_3+l_4$.  We wish to replace these four variables by three variables that can be integrated from zero to infinity independently, as in Eq.\ (\ref{eq:reexpress-momenta}).  

 In Eq.\ (\ref{eq:E-integrals-loops}), we have for Fig.\ \ref{fig:4-point-example},
 \bea 
  G_{(2,2)\P}\left (\{x_4,x_3\}_{\rm out},\{x_2,x_1\}_{\rm in}\right ) & = & (2\pi)\, \frac{(-ig)}{(4\pi^2)^4}\,  \int  d^3y  \prod_{k=1}^4  \theta(z_k^+)\  
  \prod_{j=3}^4 \int_0^\infty dl_j\, \prod_{i=1}^2 \int_0^\infty dk_i\, 
  \nn\\
  &\ & \hspace{10mm} \times\ \delta \left ( l_3+l_4-k_1-k_2  \right )
  \nn\\
&\ & \hspace{10mm} \times
  \prod_{j=3}^4
  \frac{\theta \left ( l_j  \right ) }{2(x_j^+-y^+)}\, 
   \exp\left(-il_j \left (x^-_j -  \frac{(x_{j\perp}-y_{\perp})^2+i\ep}{2(x_j^+-y^+)} \right ) \right) 
  \nonumber\\
  &\ & \hspace{10mm} \times\ 
  \prod_{i=1}^2 
  \frac{\theta \left ( k_i \right ) }{2(y^+ - x_i^+)}\, 
  \exp\left(-ik_i \left (- x^-_i- \frac{(x_{i\perp}-y_\perp)^2+i\ep}{2(x_i^+-y^+)} \right ) \right) \, .
  \label{eq:E-integrals-2-to-2}
  \eea
 We denote the three independent integration variables that we seek as $q_{dc}$, and to cover the full integration space of the light-cone energy conserving $k_i$ and $l_j$, we need two sets of corresponding variables and paths.       One complete choice ${\cal C} =\{\pi_1,\pi_2\}$ 
 for the sets of paths $(dc)$, and equivalently the light-cone energies $q_{dc}$, which cover the constrained space of $k_j$ and $l_i$ integrals is given by
  \bea
 \pi_1\ &=&\ \{ (31),(32),(42) \} 
  \nonumber\\
\pi_2\ &=&\ \{  (31),(41),(42) \}\, .
  \label{eq:pi-choice}
\eea
These sets correspond simply to the two possibilities:  $l_3>k_1$ and $l_3<k_1$, respectively.
In Eq.\ (\ref{eq:reexpress-momenta}), the coverings $\pi_1$ and $\pi_2$  are defined by the integration variables 
\begin{align}
\hspace{-15mm} \pi_1: \   k_1\ &=\ q_{31} \hspace{15mm} &\pi_2: \hspace{5mm} &k_1=q_{31} + q_{41} \nn\\ 
k_2\ &=\ q_{32}+q_{42}  & &k_2=q_{42} \nn\\
l_3\ &=\ q_{31}+q_{32}\,  & & l_3=q_{31} \nn\\
l_4\ &=\ q_{42}\, ,  & & l_4=q_{41}+ q_{42}\, ,
\label{eq:reexpress-momenta-4point}
\end{align}
with each of the $q_{dc}$ integrated from zero in infinity in both cases.   The integrals are now trivial, leading to three path denominators, as in the examples developed above, but now resulting directly from the $q_{dc}$ integrals.   We note, however, that we have made a choice in partitioning the integration space according to Eq.\ (\ref{eq:pi-choice}).   We could equally well have reversed the roles of the final vertices $x_3$ and $x_4$, or equivalently the light-cone energies $l_3$ and $l_4$.   

Naturally, these two choices must give the same answer, because they are related by a change of variables, but only after adding together the terms associated with the two choices of paths in each case.   Explicitly, we have
\bea
&\ & \hspace{-10mm} i^3\, \int_0^\infty d q_{31}\ d q_{32}\ dq_{42}\ 
e^{-iq_{31} \left( x^-_3-x^-_1-D_{31}-i\ep\right) }
e^{-iq_{41} \left( x^-_3-x^-_1-D_{32}-i\ep\right) }
e^{-iq_{42} \left( x^-_4-x^-_2-D_{42}-i\ep\right) }
\nn\\
&\ & \hspace{-5mm} +\ i^3\,  \int_0^\infty d q_{31}\ d q_{41}\ dq_{42}\ 
e^{-iq_{31} \left( x^-_3-x^-_1-D_{31}-i\ep\right) }
e^{-iq_{41} \left( x^-_4-x^-_1-D_{41}-i\ep\right) }
e^{-iq_{42} \left( x^-_4-x^-_2-D_{42}-i\ep\right) }
\nn\\
&\ &\ \hspace{-10mm} =\ \frac{1}{x^-_3-x^-_1-D_{31}-i\ep}\frac{1}{x^-_3-x^-_2-D_{32}-i\ep} \frac{1}{x^-_4-x^-_2-D_{42}-i\ep}
\nn\\
&\ & \hspace{-5mm}
\ +\ \frac{1}{x^-_3-x^-_1-D_{31}-i\ep}\frac{1}{x^-_4-x^-_1-D_{41}-i\ep} \frac{1}{x^-_4-x^-_2-D_{42}-i\ep}
\nn\\
 &\ & \hspace{-10mm}\ =\ \frac{1}{x^-_4-x^-_1-D_{41}-i\ep}\frac{1}{x^-_4-x^-_2-D_{42}-i\ep} \frac{1}{x^-_3-x^-_2-D_{32}-i\ep}
\nn\\
 &\ & \hspace{-5mm}\  +\ \frac{1}{x^-_4-x^-_1-D_{41}-i\ep}\frac{1}{x^-_3-x^-_1-D_{31}-i\ep} \frac{1}{x^-_3-x^-_2-D_{32}-i\ep}
\nn\\
&\ & \hspace{-10mm}\ =\ i^3\,  \int_0^\infty d q_{41}\ d q_{42}\ dq_{32}\ 
e^{-iq_{41} \left( x^-_4-x^-_1-D_{41}-i\ep\right) }
e^{-iq_{42} \left( x^-_4-x^-_2-D_{42}-i\ep\right) }
e^{-iq_{32} \left( x^-_3-x^-_2-D_{32}-i\ep\right) }
\nn\\
&\ & \hspace{-5mm} 
+\ i^3\,  \int_0^\infty d q_{41}\ d q_{31}\ dq_{32}\ 
e^{-iq_{41} \left( x^-_4-x^-_1-D_{41}-i\ep\right) }
e^{-iq_{31} \left( x^-_3-x^-_1-D_{31}-i\ep\right) }
e^{-iq_{32} \left( x^-_3-x^-_2-D_{32}-i\ep\right) }\, ,
\nn\\
\label{eq:4pt-identity}
\eea
with the $D_{ba}$ defined as in (\ref{eq:1st-deficit-def}).
The equality, as represented in the middle of (\ref{eq:4pt-identity}), is confirmed algebraically, directly from the identity, 
$
D_{41}-D_{42} = D_{31}-D_{32}
$\@. 
This is the pattern we will find for any ordered diagram, and at any order in perturbation theory.   We now turn to the general construction.

 \subsection{The construction}
 \label{subsec:construction}
 
We are 
ready to prove that any ordered diagram $G_{(n,m)\P}$ can be written as a sum of terms with initial-to-final paths, Eq.\ (\ref{eq:1st-result-massless}).   For a general diagram, the choice of integration variables that produce paths from initial to final vertices is not unique.  Even for a tree diagram, we may have many choices, just as there are many choices of loop momenta.  We will see, however, that all consistent choices have the same qualitative features.    In the following, we use the notation of Eq.\ (\ref{eq:k-l-notation}), and all light-cone energies flowing from initial to final vertices are positive.   We begin by specifying all light-cone energies $l_j$, $k_i$, and we want to show how to cover the entire space of these light-cone energies, subject only to overall conservation.  This will provide a change of variables in Eq.\ (\ref{eq:E-integrals-loops}) that can be integrated trivially to give path denominators like those in Eq.\ (\ref{eq:1st-result-massless}).  We begin with tree diagrams, and 
afterwards show that the construction can be extended to loop diagrams.

\subsubsection{Partitions of momentum flow}

 As we have seen, in the general expression, Eq.\ (\ref{eq:E-integrals-loops}), for an ordered diagram, light-cone momenta can flow only from precedent to descendant vertices.  We can define, then, sets, $T_j$ of initial vertices $x_i^{(j)}$,  consisting of all the predecessors of each final vertex, which we label $X_j$.  Following our 
  notation above, the indices labelling final vertices run from $j=m+1,\dots , m+n$.  We provide an arbitrary ordering to the elements of this set,
 \bea
 T_j\ \equiv\ \left \{ x^{(j)}_{1},x^{(j)}_2, \dots, x^{(j)}_{\kappa_j}\  |\ X_j\ >\  x_i^{(j)}\right \}\, ,
 \eea
  where $\kappa_j$ is the number of elements in the set.  In the general case each initial vertex $x_i$ will appear in several such sets, $T_j$.
 We will identify every covering set of paths $\pi_I$ as a map from the $n$ sets $T_j$ to another group of $n$  sets, $S_j$,
 \bea
 \pi_I\ : \{ T_j \} \ \rightarrow\ \{S_j\}\, .
 \label{eq:flow-map}
 \eea
 As a map, $\pi_I$ defines a set of $m+n-1$ integration variables that specify the change of variables in Eq.\ (\ref{eq:reexpress-momenta}).   Specifically,  if $x^{(j)}_i$ is an element of $S_j$,  then  a light-cone-energy variable  
  $q_{ji}$  that flows from initial vertex $x^{(j)}_i$ to final vertex $X_j$ will be chosen as an independent integration variable as in Eq.\ (\ref{eq:reexpress-momenta}), to be integrated from zero to infinity.
 If we denote the number of elements in set $S_j$ by $\nu_j$, then we will find that
 $
 \sum_{j=m+1}^{m+n} \nu_j\ =\ m+n-1\, ,
 $
where $\nu_j\le \kappa_j$.   We will give a prescription for producing the $S_j$'s out of the $T_j$'s for connected diagrams.   This construction will be unique, once we have identified the sets $T_j$, $j=m+1, \dots, m+n$, provided each with an ordering, and have assigned values to all incoming and outgoing momenta $k_i$ and $l_j$.   In essence the construction will consist of a prescription for discarding elements of the $T_j$'s in a manner that depends on the values of the $k$'s and $l$'s.
 
  The covering sets, $\pi_I$ are generated by a simple set of steps, the nature of which assures that the partition is exhaustive in $l_j/k_i$ space, and that the subspaces corresponding to these partitions are nonoverlapping.   
 The method is a straightforward extension of the motivation used to identify the sets $\pi_1$ and $\pi_2$ in the analysis of the covering set for the trivial diagram, Eq.\ (\ref{eq:reexpress-momenta-4point}).   
 
 We start with set $T_{m+1}$, dealing first with  light-cone energy $l_{m+1}$, the momentum associated with final vertex $X_{m+1}$,  and $k_1$, the light-cone energy flowing out of vertex $x^{(m+1)}_1 \in T_{m+1}$.     By construction, initial vertex $x^{(m+1)}_1$ is a predecessor of final vertex $X_{m+1}$.

For the pair  $l_{m+1}$ and $k_1$, we have, of course, either $l_{m+1} > k_1$ or $l_{m+1}< k_1$.  The equality may be associated with either case.  We then proceed as follows.

\begin{itemize}

\item In the range $l_{m+1} > k_1$, we define $q_{m+1\,1}\equiv k_1$ to flow from initial vertex $x^{(m+1)}_1$ to final vertex $X_{m+1}$, and then consider the momentum flow in the remaining diagram.  By definition there is a forward-moving path between vertex  $x^{(m+1)}_1$ and $X_{m+1}$, and $q_{m+1\, 1}$ is chosen to flow along that path.    In a tree diagram, the path is unique, since two paths would define a loop.   The flow of light-cone energy through the diagram from vertex $x^{(m+1)}_1$ is now fixed.   This effectively eliminates one initial vertex from the problem, which is recast with $m-1$ initial vertices and $n$ final vertices, with the same momentum flow as previously, except that now light-cone energy 
\bea
l_{m+1}'\ =\ l_{m+1}-q_{m+1\,1}>0
\eea
 flows into the final vertex $X_{m+1}$ in the modified flow.  For the example of Fig.\ \ref{fig:4pexA}, this possibility corresponds to the momentum flow $\pi_1$ in Eq.\ (\ref{eq:reexpress-momenta-4point}), and $q_{31}$ is integrated freely from zero to infinity.   In the general expression, Eq.\ (\ref{eq:E-integrals-loops}), the $k_1(=q_{m+1\, 1})$ integral can be performed to give the denominator corresponding to the path from $x_1^{(m+1)}$ to $X_{m+1}$, while the remaining $m+n-1$ integrals are unchanged, except that $l_{m+1}$ is everywhere replaced by $l_{m+1}'$, including in the argument of the light-cone-energy conservation delta function.

Note that, in general, if $l_{m+1} > k_1$,  vertex $X_{m+1}$ must be a descendant of at least one other initial vertex.   Among these initial vertices, at least one must be the source of the additional light-cone energy that appears on $l_{m+1}$.   We pick the second initial vertex $x^{(m+1)}_2$ in the ordering of $T_{m+1}$, which by construction is also a predecessor of $X_{m+1}$.   Because vertex $X_{m+1}$ also receives energy from (at least) vertex $x^{(m+1)}_2$, our map (\ref{eq:flow-map}) becomes
\bea
T_{m+1}\ \rightarrow\ S_{m+1}\ =\ \{ x^{(m+1)}_1,x^{(m+1)}_2, \dots\}\, .
\eea
For the example of Fig.\ \ref{fig:4pexA}, $T_3=T_4=\{x_1,x_2\}$, while $S_3=\{x_1,x_2\}$ and $S_4=\{x_2\}$.

\item For the second, case, $l_{m+1}<k_1$, we define
\bea
q_{m+1\,1}\ &=&\ l_{m+1}\, ,
\nn\\
k_1'\ &=&\ k_1\ -\ l_{m+1}\, ,
\eea
where again, in a tree diagram there is a unique light-cone energy flow from initial vertex $x_1^{(m+1)}$ to final vertex $X_{m+1}$.   This procedure effectively removes final vertex $X_{m+1}$ from the problem, which is recast as $n-1$ final and $m$ initial vertices, with the same light-cone energy flows as previously, except that now $k_1$ is replaced by $k_1'$.   
In the general expression, Eq.\ (\ref{eq:E-integrals-loops}), it is now the $l_1(=q_{m+1\, 1})$ integral that can be performed to give a denominator corresponding to the path from $x_1^{(m+1)}$ to $X_{m+1}$, while the remaining $m+n-1$ integrals are unchanged, except that now $k_1$ is everywhere replaced by $k_1'$, including in the argument of the light-cone-energy conservation delta function.
 For this case, the first step defines
\bea
T_{m+1}\ \rightarrow\ S_{m+1}\ =\ \{ x^{(m+1)}_1 \}\, ,
\label{eq:pi-as-map}
\eea
The relevant example above is Fig.\ \ref{fig:4pexB}, with momentum flow $\pi_2$ in Eq.\ (\ref{eq:reexpress-momenta-4point}), and $l_3\equiv q_{31}$. 
Thus, for the example of Fig.\ \ref{fig:4pexB}, $S_3=\{x_1\}$ and  $S_4=\{x_1,x_2\}$.

\end{itemize}

Repeating this process, at each step we eliminate either one final or one initial vertex from the problem, identifying a single momentum integral from zero to infinity, while uniquely assigning light-cone energies for the remaining external vertices.   Eventually, we will arrive at an energy flow with only one initial and/or one final vertex. At this point there is only a single free integral left, leading to a total of $m+n-1$ free integrals in place of the original set of $m+n$ integrals constrained by momentum conservation.   Note that if there is only one initial vertex ($m=1$) and any number of final vertices, then the identity $k_1>l_j$ will hold for all final state $l_j$, $j=2,\dots ,1+n$, and the number of final vertices will be decreased at each step in the process.   Similarly,  if there is only a single final vertex and some number of initial vertices, the number of initial vertices decreases at each step.   In the last step, the delta function of Eq.\ (\ref{eq:E-integrals-loops}) will reduce the final two remaining integrals to a single integral, which flows between the remaining two vertices, providing the last path denominator.  The total number of identified integrals and denominators is then $m+n-1$.   The result of this process is a list ${\cal C}$, a class of  maps 
$\pi_I$, covering sets that map the sets of predecessor initial and descendant final vertices into pairs, as in Eq.\ (\ref{eq:pi-as-map}).   
As we have seen in an example and shown above, these pairings are determined once an arbitrary ordering has been imposed on the initial and final vertices at the start of the construction.

\begin{figure}[b]
\centering
\subfigure[]{
\includegraphics[height=3.5cm]{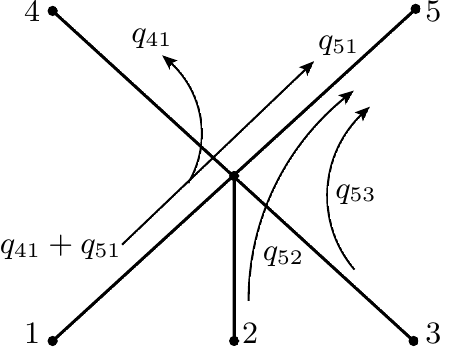}\label{fig:3to2A}}
\hspace{10mm}
\subfigure[]{
\includegraphics[height=3.5cm]{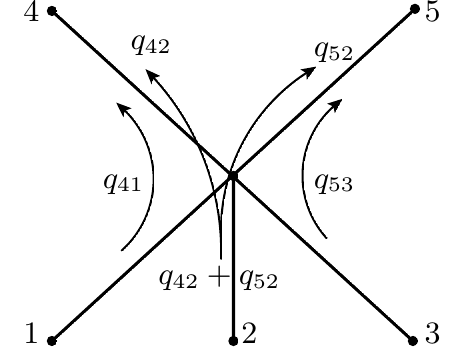}\label{fig:3to2B}}
\hspace{10mm}
\subfigure[]{
\includegraphics[height=3.5cm]{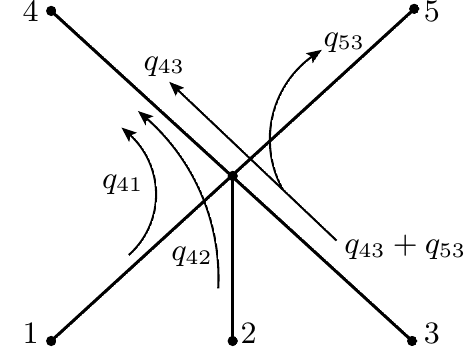}\label{fig:3to2C}}

\caption{An example of light-cone-energy flows for a three-to-two process, for \subref{fig:3to2A} $l_4<k_1$, \subref{fig:3to2B} $k_1<l_4<k_1+k_2$, and \subref{fig:3to2C} $k_1+k_2<l_4$, as discussed in the text.  \label{fig:3-to-2}}

\end{figure}

As an illustration slightly beyond the simple example of the previous subsection,  we consider the momentum flows for a diagram with five external lines attached  by a single vertex, for the relative ordering illustrated in Fig.\ \ref{fig:3-to-2}.   Each of the two final vertices is a descendant of all three initial vertices, so that here $T_j=\{x_1,x_2,x_3\}$, $j=4,5$.   Following our construction, we find three coverings, which can be identified from the value of the final light-cone energy $l_4$ relative to the initial light-cone energies, $k_1$, $k_2$, and $k_3$.  We will always find four light-cone momenta $q_{fi}$ to replace the five constrained integrals over the $k$'s and $l$'s.

Applying the construction above, the coverings are identified through inequalities, in this case involving $l_4$.   Covering $\pi_1$ corresponds to $l_4<k_1$, so that $S_4=\{x_1\}$ and $S_5=\{x_1,x_2,x_3\}$.   In the notation of Eq.\ (\ref{eq:reexpress-momenta}), covering $\pi_1$ has  integration variables $q_{41}$, $q_{51}$, $q_{52}$, and $q_{53}$.  Covering $\pi_1$ is represented by Fig.\ \ref{fig:3to2A}.  The next covering, $\pi_2$,  is defined by $k_1<l_4<k_1+k_2$, with integration variables $q_{41},q_{42},q_{52},q_{53}$, corresponding to $S_4=\{x_1,x_2\}$ and $S_5=\{x_2,x_3\}$, Fig.\ \ref{fig:3to2B}.   In the remaining region, $l_4>k_1+k_2$, with variables $q_{41},q_{42},q_{43},q_{53}$, with $S_4=\{x_1,x_2,x_3\}$ and $S_5=\{x_3\}$, Fig.\ \ref{fig:3to2C}.  As in the example of Eqs.\ (\ref{eq:reexpress-momenta-4point}) and (\ref{eq:4pt-identity}) above, permutations of the initial or final vertices lead to equivalent results.

In summary, because the construction described here exists for any choice of light-cone energies flowing through the initial and final vertices, and because it proceeds at each step through inequalities in the energy flow of  individual initial and final vertices, the terms that are generated by the maps $\pi_I$ provide a complete, disjoint covering of the full constrained $k_i/l_j$ space, replacing it with a sum of terms in which each integral is associated uniquely with an initial-to-final path and is freely integrated from zero to infinity.

\subsubsection{Including loops}

For tree diagrams, the construction above determines the light-cone energy of each line.   For ordered diagrams with loops, on the other hand, we cannot depend on having a unique path between pairs of initial and final vertices, and our procedure does not yet fix the light-cone energy of every line.  We can, however, use the same construction to treat loop diagrams, by analyzing tree diagrams that are produced by cutting lines within 
one loop diagrams, and then proceeding inductively.

 We assume that up to $K-1$ loops, we have already shown that any ordered diagram can be written as a sum of terms, each with $E+K-1$ independent integrals over forward-moving, positive light-cone energies.   For a $y^+$-ordered diagram with $E$ external lines and $K$ loops, we begin by cutting any line that reduces the loop diagram to $K-1$ loops.   Because the light-cone energy flow in any line must be toward larger $y^+$, we may treat the cut line as a pair of external lines, one of which flows into the diagram from a new initial vertex, and one of which flows out of the diagram to a new final vertex, treating them as two additional light-cone energies.   Now the diagram has $E+2$ external lines.  
 
\begin{figure}[b]
\centering
\includegraphics[height=3.6cm]{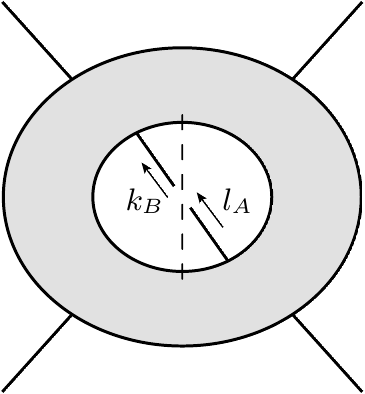}

\caption{A general diagram with a cut line, discussed in connection with light-cone energy flow through loop diagrams.  In this argument, $k_B$ is treated as an initial energy and $l_B$ as a final energy, subject to the constraint $l_A=k_B$.   \label{fig:cut-line-fig}}

\end{figure}

The generic form is illustrated by Fig.\ \ref{fig:cut-line-fig}, where $p \equiv k_B \equiv l_A$ is the momentum of the cut line.   Momentum $k_B$ is to be treated as one of the initial momenta of the diagram cut in this way, and $l_A$ as one of the final momenta.    Performing the same analysis as above for the resulting diagram, we derive a sum of terms in which each of these $(E+K-1)+2$ energies is given as a linear combination of a new set of $E+K$ positive incoming and outgoing light-cone energies, as in Eq.\ (\ref{eq:reexpress-momenta}).     Because we assume the result at $K-1$ loops, $k_B$ and $l_A$ can be expressed in the form
\bea
k_B\ &=&\ \sum_i q_{iB}\, ,
\nn\\
l_A\ &=&\ \sum_j q_{Aj}\, ,
\eea
where $q_{iB}$ and $q_{Aj}$ make up a subset of the complete set of light-cone energies for the cut diagram with $k_B$ and $l_A$ treated as independent (note that each of these sums might have only one term).   The original diagram is found by integrating over $k_B$ and $l_A$ with an energy constraint, which in the new variables becomes
\bea
\prod_{\{iB\},\{Aj\}} \int dq_{iB}\; dq_{Aj} \ \delta\left(\sum_i q_{iB}\ -\ \sum_j q_{Aj} \right)\, .
\eea
 We recognize here  the same problem of constrained integration, with the same solution in terms of a sum over changes to variables that can be integrated freely from zero to infinity.    We may therefore treat this constraint by the same method as above to the $n_p$ independent energies that flow into $k_B$ and out of $l_A$.   As above, this produces a sum of terms in which the number of independent, positive external light-cone energies is reduced to $n_p-1$,
 \bea
 \left\{ {\{q_{iB} \}, \{q_{Aj} \}} \right\}\ \rightarrow \left\{ q'_{f(BA)i}\right \}\, ,
 \eea
 where the new variables $q'_{f(BA)i}$ have unconstrained integrals from zero to infinity.   By construction, they all flow from initial vertices of the original diagram, through the cut line, $p$, to final vertices through forward-moving paths.

   The final result is a sum of terms in which the number of independent positive energies is reduced by one, from $E+K$ to $E+K-1$, leaving us with the number of integrations for a $K$-loop diagram with $E$ external lines.   By construction, each of these momenta will flow from an initial to a final vertex.

\subsection{Summary: the general result}
\label{sec:general-construction}

For any ordered diagram $G_{\P}$, the construction above provides a ``covering class", that is, a set ${\cal C}=\{\pi_I\}$, each element of which is a 
covering set of paths that include every line in the diagram, the sum of whose allowed values of light-cone energies covers the entire integration region in the corresponding integral, Eq.\ (\ref{eq:E-integrals-loops}).   As we have seen, the choices of these sets of paths are not unique, because the construction involves an arbitrary ordering of initial and final vertices,   but any class generated in this way is equivalent.   The integral in terms of any specific covering class ${\cal C}$ is then
 \bea 
  G_{(n,m)\P}\left (\{x_d\}_{\rm out},\{x_c\}_{\rm in}\right )\ &=&\   \sum_{\pi_I \in {\cal C}} G_{(n,m)\P}^{(\pi_I)}\left (\{x_d\}_{\rm out},\{x_c\}_{\rm in}\right )\, ,
\eea
where the term associated with each covering $\pi_I$ is
   \bea 
  G_{(n,m)\P}^{(\pi_I)} \left (\{x_d\}_{\rm out},\{x_c\}_{\rm in}\right ) & = & (2\pi)^N\, \frac{(-ig)^N}{(4\pi^2)^L}\, (-i)^L\int\left(\prod_{i\in N} d^3y_i\right) 
  \prod_{j=1}^{m+n+K-1} \int_0^\infty dq^{(\pi_I)}_j\, \
  \nn\\
&\ & \hspace{2mm} \times
  \prod_{w=1}^L
  \frac{\theta(z_w^+)\theta \left ( q^{(\pi_I)}_j\sigma^{(\P,\pi_I)}_{jw} \right ) }{2z^+_w}\, 
  \exp\left(-iq^{(\pi_I)}_j\sigma^{(\P,\pi_I)}_{jw} \left (\eta'_{wa}{}^{(\P)}x^-_a-\frac{z^2_{w\perp}+i\ep}{2z^+_w}\right ) \right) \, ,
  \label{eq:GP-covering}
  \eea
 where the incidence matrix $\sigma^{(\P,\pi_I)}_{jw}=+1$ if light-cone energy $q^{(\pi_I)}_j$ flows on line $w$, or is zero otherwise.  The theta functions are therefore all unity over the entire integration region, and we can now do the integrals over the new variables $q^{(\pi_I)}_j$. 

 Our final result, as proposed in Eq.~(\ref{eq:1st-result-massless}) but now specifically for each covering set $\pi_I$, is given by
  \bea
G_{(n,m)\P}^{(\pi_I)}\left (\{x_d\}_{\rm out},\{x_c\}_{\rm in}\right )\ & = & (2\pi )^N\ \frac{(-g)^N}{(4\pi^2)^L}\, \int\left(\prod_{i\in N} d^3y_i\right) \prod_{{\rm all}\ j} \frac{\theta(z_j^+)}{2z_j^+}\  
 \prod_{\{P_{(ba)}^{(\pi_I)}\}}\   \frac{-1}{x^-_b\ -\ x^-_a\ -\ D^{(\pi_I)}_{(ba)} -i\ep }\, ,
 \label{eq:result-massless}
\eea 
where each $P_{(ba)}^{(\pi_I)}$ denotes a path  in the covering set $\pi_I$, extending from some initial vertex $x_a\in \{x_c\}_{\rm in}$ to some final vertex $x_b\in \{x_d\}_{\rm out}$, and where $D^{(\pi_I)}_{(ba)}$ is the ``light-cone distance" associated with the path, defined by the plus and transverse components of the intermediate vertices, 
as in Eq.\ (\ref{eq:1st-deficit-def}),
\bea
D^{(\pi_I)}_{(ba)}\ =\ \sum_{i\in P_{(ba)}^{(\pi_I)}} \frac{z_{i,i-1,\perp}^2}{2z_{i,i-1}^+}\, .
\label{eq:deficit-def}
\eea
Together, Eqs.\  (\ref{eq:result-massless}) and (\ref{eq:deficit-def}) specify the expression of an arbitrary diagram with massless scalar propagators in terms of path denominators.  This is the coordinate analog of the momentum-space expression, Eq.\ (\ref{eq:lcopt-mtm-basic}).   In the appendix, we provide an alternate proof of this result, based on the direct evaluation of integrals over minus components of internal vertices in ordered diagrams.

In the next sections, we discuss some of the properties of these expressions, including 
how imaginary parts and discontinuities arise, and how to derive the eikonal approximation in coordinate space.   We will then be ready to apply our analysis to expectations of Wilson lines,
and derive some of the interesting results found in Refs.\ \cite{Laenen:2014jga,Laenen:2015jia} from our general viewpoint.
We go on to discuss how to treat factors associated with spin, and finally how to use dimensional regularization and incorporate masses.

\section{The Coordinate Eikonal Approximation and Wilson Lines}
\label{sec:eikonal-Wilson}

So far, our discussion has treated purely scalar theories.  In the following, we show that many of the qualitative features of scalar theories extend to arbitrary perturbative expansions with numerator factors, including gauge theories.   This becomes evident by returning to the triangle diagram, this time in the context of scalar quantum electrodynamics.   We will see that the basic light-cone coordinate integrals lead to results that are equivalent to those found in the scalar case, but supplemented by derivatives for certain terms.   We will also show that such terms lead  to the eikonal approximation in the limit that external lines approach the light cone.   The eikonal approximation then leads us naturally to the coordinate representation for Wilson lines.  

\subsection{Vertex for scalar QED}

\begin{figure}

\centering
\includegraphics[height=3.5cm]{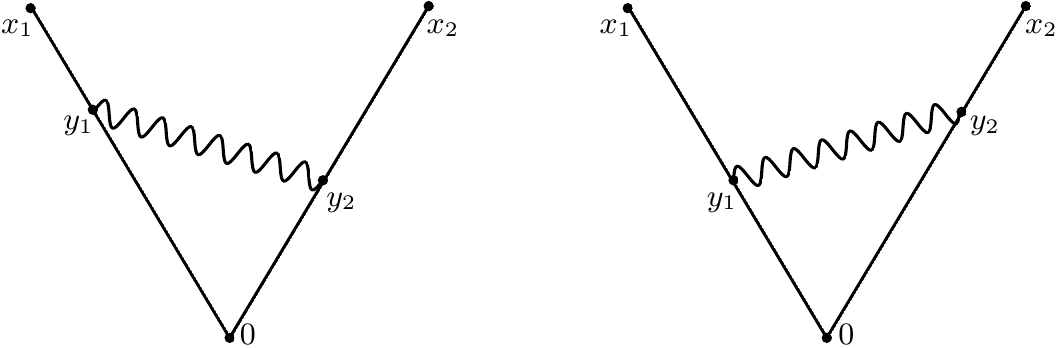}

\caption{Two orderings for a charged scalar vertex with $x_{1,2}^+>0$. \label{fig:chared-scalar}}

\end{figure}

We now discuss in some detail the example of the ordered diagrams in Fig.\ \ref{fig:chared-scalar}, the charged scalar vertex correction in coordinate space, 
\bea
\Gamma(x_1,x_2) & = & \left(\frac{1}{4\pi^2}\right)^5(-ie)^2\int d^4y_1\,d^4y_2\left\{\left(\partial_{y_2}^{\mu}\frac{1}{-(x_2-y_2)^2+i\ep}\right)\frac{1}{-y^2_2+i\ep} - \frac{1}{-(x_2-y_2)^2+i\ep}\left(\partial_{y_2}^{\mu}\frac{1}{-y^2_2+i\ep}\right) \right\} \nn \\
 &\  & \quad\times \left[  \frac{-g_{\mu\nu}}{-(y_2-y_1)^2+i\ep}\right]\left\{\left(\partial_{y_1}^{\nu}\frac{1}{-(x_1-y_1)^2+i\ep}\right)\frac{1}{-y^2_1+i\ep} - \frac{1}{-(x_1-y_1)^2+i\ep}\left(\partial_{y_1}^{\nu}\frac{1}{-y^2_1+i\ep}\right) \right\} \nn \\
 & = &  \left(\frac{1}{4\pi^2}\right)^5\,4e^2\int d^4y_1\,d^4y_2\left\{\frac{(y_2-x_2)^{\mu}}{[-(x_2-y_2)^2+i\ep]^2}\frac{1}{-y^2_2+i\ep} - \frac{1}{-(x_2-y_2)^2+i\ep}\frac{y^{\mu}_2}{[-y^2_2+i\ep]^2} \right\} \nn \\
 &\  & \quad\times \frac{g_{\mu\nu}}{-(y_2-y_1)^2+i\ep}\left\{\frac{(y_1-x_1)^{\nu}}{[-(x_1-y_1)^2+i\ep]^2}\frac{1}{-y^2_1+i\ep} - \frac{1}{-(x_1-y_1)^2+i\ep}\frac{y_1^{\nu}}{[-y^2_1+i\ep]^2} \right\} \ . 
 \label{eq:qed-Gamma}
\eea
Compared to the diagrams in Fig.\ \ref{fig:ordered-scalar-tri}, we   fix the vertex that creates a pair of charged scalars at the origin as the earliest vertex (with $x_{1,2}^+>0$ for the other external vertices), and evaluate the resulting diagram. 
We now show how to modify the analysis above, to obtain the QED analog of the ``path denominator" forms, Eqs.\ (\ref{eq:path-form-example-5plus}) and (\ref{eq:path-form-example-5minus}) for the two orderings in Fig.\ \ref{fig:ordered-scalar-tri}, in the presence of the numerator factors.   To carry this out, we will show that the numerators can be taken into account entirely in terms of plus and transverse integration variables.

For an arbitrary line with coordinate distance $w^\mu$, we introduce 
a lightlike auxiliary vector,
\bea
\hat w^\mu \ =\ \left( w^+ \, , \, \frac{w_{\perp}^2}{2w^+}\, , \, w_{\perp} \right)\, ,  \quad\ \quad \hat w^2=0 \ .
\label{eq:hatw-def}
\eea
We now follow the steps leading from the first to second lines of Eq.\ (\ref{eq:delta-to-E}) to
 rewrite  propagtors with numerator factors, as in Eq.\ (\ref{eq:qed-Gamma}),  in integral form as
\bea
\frac{w^\mu}{(-w^2\ +\ i\ep)^2} & = & \frac{1}{2w^+}\frac{\partial}{\partial \hat w^-}\
\left[ \,\hat w^{\mu} \ \int_{-\infty}^\infty dE^+\left\{   \frac{i}{2|w^+|} \theta(E^+w^+)\, e^{iE^+(\hat w^- -\, w^- +\, i\ep/2w^+)}\right\} \right ]
\nn\\
 & = & \frac{1}{2w^+}\frac{\partial}{\partial \hat w^-}\ \left[  \frac{-\ \hat w^{\mu}} {-2w^+\left(w^-\ -\ \hat w^- \right) + \, i\ep} \right]\ ,
\label{eq:hatw-deriv}
\eea
where by (\ref{eq:hatw-def}) the vector $\hat w^\mu$ is lightlike but shares plus and transverse components with $w^\mu$.   This result is easy to check by direct evaluation of either right-hand side. The partial derivative with respect to $\hat w^-$ is carried out at fixed $w_\perp$ and $w^+$, and in a general expression acts only on the propagator of a specific line.
With this result, quite generally, all integrals in theories with spin and momentum factors at vertices can be written in terms of scalar integrands acted on by operations that commute with the minus integrations.   Thus, we may use results of scalar integrals first.

In these terms, we write the vertex function as
\bea
\Gamma(x_1,x_2) & = & \left(\frac{1}{4\pi^2}\right)^5\,4e^2\int d^3y_1\,d^3y_2  \left[ \frac{1}{2z^+_{11}}\frac{\partial}{\partial \hat z^-_{11}}\frac{1}{2z^+_{22}}\frac{\partial}{\partial \hat z^-_{22}}\,\hat z_{11}\cdot \hat z_{22} \ +\  \frac{1}{2z^+_{11}}\frac{\partial}{\partial \hat z^-_{11}}\frac{1}{2z^+_{2}}\frac{\partial}{\partial \hat z^-_{2}}\,\hat z_{11}\cdot \hat z_{2}\right. \nn \\
 & \ & \hspace{2cm} +\ \left. \frac{1}{2z^+_{1}}\frac{\partial}{\partial \hat z^-_{1}}\frac{1}{2z^+_{22}}\frac{\partial}{\partial \hat z^-_{22}}\,\hat z_{1}\cdot \hat z_{22}\ +\ \frac{1}{2z^+_{1}}\frac{\partial}{\partial \hat z^-_{1}}\frac{1}{2z^+_{2}}\frac{\partial}{\partial \hat z^-_{2}}\,\hat z_{1}\cdot \hat z_{2}  \right]\,\gamma_{\rm scalar}(z_i,z_{ii})  \ ,  
 \label{eq:scalar-vertex-hatz}
\eea
where we have defined
\bea
z_i\ &\equiv& \  y_i
\nn\\
z_{ii}\ &\equiv& \ x_i\ -\ y_i
\nn\\
z_{12}\ &\equiv&\ y_1\ -\ y_2\, .
\eea
The function $\gamma_{\rm scalar}$ in Eq.\ (\ref{eq:scalar-vertex-hatz}) includes integrals over the $y_i^-$ variables, using the same steps as in the scalar case, introducing light-cone energy variables, which may themselves be integrated to give path denominators.

As in the scalar case, we choose $x_1^+,x_2^+>0$, so that there are only two possible partial orderings, corresponding to the results of Eq.\ (\ref{eq:path-form-example-5plus}) and (\ref{eq:path-form-example-5minus}).
We show explicitly the case $y_1^+>y_2^+$, which we identify by a superscript,  
\bea
\gamma^{(12)}_{\rm scalar} & = &  \left(\frac{-i \prod_I^{(12)} \theta\left( z^+_I\right)}{2^5\prod_I z^+_I}\right) \ \int^{\infty}_0 \frac{dE^+_1}{2\pi}\frac{dE^+_{2}}{2\pi}\frac{dE^+_{11}}{2\pi}\frac{dE^+_{22}}{2\pi}\frac{dE^+_{12}}{2\pi}\,(2\pi)\delta(E^+_{11}-E^+_{1}-E^+_{12}) \nn \\
 & \ & \hspace{1.2cm}\times\ (2\pi)\delta(E^+_{2}-E^+_{22}-E^+_{12})\times e^{iE^+_{12}\hat z^-_{12}+iE^+_{11}(\hat z^-_{11}-x^-_1)+iE^+_{22}(\hat z^-_{22}-x^-_2)+iE^+_1\hat z^-_1 + iE^+_2\hat z^-_2} \nn \\
& = & \ \left(\frac{-i\, \prod_I^{(12)} \theta\left( z^+_I\right)}{2^5\prod_I z^+_I}\right)  \ \int^{\infty}_0 \frac{dE^+_1}{2\pi}\frac{dE^+_{12}}{2\pi}\frac{dE^+_{22}}{2\pi} \, e^{iE^+_{12}\hat z^-_{12} +i(E^+_1+E^+_{12})(\hat z^-_{11}-x^-_1)} \nn \\
 & \ & \hspace{3.5cm} \times\ e^{iE^+_{22}(\hat z^-_{22}-x^-_2)+ iE^+_1\hat z^-_1 + i(E^+_{22}+E^+_{12})\hat z^-_2} \nn \\
 & = & i^3\,  \left(\frac{-i\, \prod^{(12)}_I \theta\left( z^+_I\right)}{2^5\prod_I z^+_I}\right) \; \frac{1}{x^-_1-\hat z^-_{11}-\hat z^-_{1}}\,\frac{1}{x^-_2-\hat z^-_{22}-\hat z^-_{2}}\,\frac{1}{x^-_1-\hat z^-_{11}-\hat z^-_{12}-\hat z^-_{2}} \ . 
 \label{eq:12-scalar}
\eea
Note that here we have fixed one vertex at the origin, and there is no ``incoming" line, so that there is one less term in the denominators. The superscript on the product over step functions indicates that it includes $\theta(y^+_1-y^+_2)$. 

\subsection{The eikonal approximation}

We now seek the most singular part of $\Gamma$, Eq.\ (\ref{eq:scalar-vertex-hatz}) as $x_1^\mu$ and $x_2^\mu$ approach the light cone, which we will identify with the eikonal approximation to the vertex function.  The singularities will arise in the integral from the regions where $y_i^\mu$ is proportional to $x_i^\mu$, $i=1,2$ \cite{Erdogan:2013bga,Erdogan:2014gha}.    The most singular contributions are found by acting with the derivatives of Eq.\ (\ref{eq:scalar-vertex-hatz}) on the denominators in Eq.\ (\ref{eq:12-scalar}) that vanish on the light cone, rather than on the numerators or on the denominator that includes $z_{12}$, 
which does not vanish in this limit.  
This gives an intermediate approximation, which we will identify by a superscript,
\bea
\Gamma^{(\rm eik(12))}(x_1,x_2) & = & \left(\frac{1}{4\pi^2}\right)^5\,4e^2\int d^3y_1\,d^3y_2 \;
\sum_{I=\{1,11\},J=\{2,22\}}
 \frac{ \hat z_{I}\cdot \hat z_{J}}{2z^+_{I} \, 2z^+_{J}}\,   
 \, \frac{\partial}{\partial \hat z^-_{I}}\frac{\partial}{\partial \hat z^-_{J}}\, \gamma_{\rm scalar}(\{z_I\}) 
 \nn \\
 & = &   \left(\frac{1}{4\pi^2}\right)^5\,4e^2
 \int_0^{x_1^+} dy_1^+\, \int_0^{x_2^+} dy_2^+\, \theta(y^+_1-y^+_2)\, 
 \int d^2y_{1\perp}\,d^2y_{2\perp} \!\!\!
\sum_{I=\{1,11\},J=\{2,22\}}
 \frac{ \hat z_{I}\cdot \hat z_{J}}{2z^+_{I} \, 2z^+_{J}}\,   
 \left(\frac{(-1) }{2^5\prod_I z^+_I}\right) 
 \nn\\
&\ & \hspace{10mm} \times\  \frac{1}{(x^-_1-\hat z^-_{11}-\hat z^-_{1})^2}\,\frac{1}{(x^-_2-\hat z^-_{22}-\hat z^-_{2})^2}\,\frac{1}{x^-_1-\hat z^-_{11}-\hat z^-_{12}-\hat z^-_{2}-i\ep} \ . 
\label{eq:Gamma_1-eik}
\eea
As the points $y_1$ and $y_2$ approach the light cone, we can do the transverse integrals explicitly, by changing variables.  We replace $y_i^+$ and $y_{i\perp}$ by variables that more closely represent paths between the origin and the points $x_i^\mu$ on the light cone,
\bea
y^+_2 & =& \lambda_2\,x^+_2 \ , \quad y_{2\perp}\ =\ \lambda_2\,x_{2\perp}+\eta_2 \ , \nonumber \\
y^+_1 & = & \lambda_1\,x^+_1 \ , \quad y_{1\perp}\ =\  \lambda_1\,x_{1\perp}+\eta_1 \ .  
\eea
For the denominator in Eq.\ (\ref{eq:Gamma_1-eik}) that depends on $y_1$, for example, this gives
\bea
x^-_1-\hat z^-_{11}-\hat z^-_{1} & = & x^-_1 - \frac{(x_{1\perp}-y_{1\perp})^2}{2(x^+_1-y^+_1)} - \frac{y^2_{1\perp}}{2y^+_1}  \nn \\
 & = & \frac{1}{2x^+_1}\left(x^2_1 - \frac{\eta^2_{1}}{(1-\lambda_1)\lambda_1}\right) \ .  
\eea
For simplicity, we take the light-cone limit from spacelike $x_i^2<0$.   The $\eta_{i}$ integrals are then
real,
\bea
\int d^2 \eta_{1} \ \frac{1}{\left(x^2_1 - \frac{\eta^2_{1}}{(1-\lambda_1)\lambda_1}-i\ep\right)^2}\ =\ \frac{\pi\, \lambda_1(1-\lambda_1)}{-x_1^2+i\ep} \ , 
\eea
and similarly for $\eta_2$.   The leading term is found by setting $\eta_1^2=\eta_2^2=0$ everywhere else in the integrand of (\ref{eq:Gamma_1-eik}).

The last denominator in Eq.\ (\ref{eq:Gamma_1-eik}) is then approximated by
\bea
x^-_1-\hat z^-_{11}-\hat z^-_{12}-\hat z^-_{2} & = & x^-_1 -\frac{[x_{1\perp}(1-\lambda_1)-\eta_{1}]^2}{2x^+_1(1-\lambda_1)}-\frac{(\lambda_1x_{1\perp}-\lambda_2x_{2\perp}+\eta_{1}-\eta_{2})^2}{2(\lambda_1 x^+_1-\lambda_2x^+_2)}-\frac{(\lambda_2x_{2\perp}+\eta_{2})}{2\lambda_2x^+_2} 
\nn\\
&\rightarrow& \frac{(\lambda_1x_1-\lambda_2x_2)^2}{2(\lambda_1x_1^+-\lambda_2x_2^+)} \ . 
\eea
We now define  ``velocity" vectors,
\bea
\beta_i^\mu \ =\ \frac{x_i^\mu}{x_i^+} \ . 
\eea
Then in the eikonal limit, all factors involving the inner product of $z_I$, $I=1,11$ and $z_J$, $J=2,22$, become
\bea
\frac{ \hat z_{I}\cdot \hat z_{J}}{2z^+_{I} \, 2z^+_{J}}\ \rightarrow \ \ \frac{1}{4}\, \beta_1\cdot \beta_2 \ , 
\eea
and we find the coordinate-space eikonal approximation for the ordering in question,
\bea
\Gamma^{(\rm eik(12))}(x_1,x_2)\ & = & \ \left(\frac{1}{4\pi^2}\right)^4\, \frac{e^2}{x_1^2x_2^2}\,  \int_0^{x_1^+} dy_1^+\, \int_0^{x_2^+} dy_2^+\, \theta(y^+_1-y^+_2)\, 
\frac{\beta_1\cdot \beta_2}{-\left( \beta_1 y_1^+\, -\, \beta_2 y_2^+\right)^2+i\ep} \, .
\eea
The companion ordering, with $y_2^+>y_1^+$, is of the same form, and, up to a prefactor involving only the $x_i^2$, their sum is just the eikonal approximation in coordinate space, as generated by products of Wilson lines \cite{Korchemskaya:1992je,Korchemsky:1992xv,Erdogan:2013bga,Erdogan:2011yc}.

\subsection{Expectations of Wilson lines}

For single-gluon exchange diagrams of Wilson lines, the relation between the coordinate light-cone ordered formalism and the coordinate-space analysis of \cite{Laenen:2014jga,Laenen:2015jia} is particularly close.   As in the example given above, in the eikonal approximation, or equivalently directly from the Wilson line definition, the ends of each gluon on a Wilson line must be treated as an ``external" vertex in the analysis above.   For gluon exchange diagrams without internal vertices, then, the steps between the invariant and light-cone forms collapse to simply factoring out the plus component of the distance between the two lines. The result separates into two terms for each gluon, which order the plus components of the vertices to which the gluon connects. To be specific, consider a gluon attached between two vertices at points $y_i=\lambda_i\beta_i$, $i=1,2$.   The light-cone form of the propagator is found by reexpressing it in a manner that reflects the relative values of the plus coordinates of the vertices to which it connects,
\bea
\frac{1}{-( y_2\ -\ y_1)^2+i\ep}\ &=&\ \frac{1}{-( \lambda_2\beta_2\ -\lambda_1\beta_1)^2+i\ep}
\nn\\
&=& \frac{\theta\left( \lambda_2\beta_2^+\ -\lambda_1\beta_1^+\right)}{2\left( \lambda_2\beta_2^+\ -\lambda_1\beta_1^+\right) }\left(
 \frac{1}{-\left( \lambda_2\beta_2^-\ -\lambda_1\beta_1^- \right)\ +\ \frac{ \left( \lambda_2\beta_{2\perp} - \lambda_1\beta_{1\perp} \right)^2}{2\left( \lambda_2\beta_2^+\ -\lambda_1\beta_1^+\right) } + i\ep}\right)
 \nn\\
 &\ & +\ 
  \frac{\theta\left( \lambda_1\beta_1^+\ -\lambda_2\beta_2^+\right)}{2\left( \lambda_1\beta_1^+\ -\lambda_2\beta_2^+\right) } \left(
 \frac{1}{-\left( \lambda_1\beta_1^-\ -\lambda_2\beta_2^- \right)\ +\ \frac{ \left( \lambda_1\beta_{1\perp} - \lambda_2\beta_{2\perp} \right)^2}{2\left( \lambda_1\beta_1^+\ -\lambda_2\beta_2^+\right) } + i\ep}\right) \ . 
 \label{eq:ladder-equivalence}
\eea
The second equality is precisely the light-cone ordered form.   

This simple example generalizes in a straightforward method to arbitrary combinations of ladder diagrams \cite{Falcioni:2014pka}, which are related to covariant expressions algebraically, since there are no internal light-cone coordinates to be integrated.   We will encounter an example with an internal vertex and corresponding integration in our discussion of discontinuities in Green functions and expectations of Wilson lines, to which we now turn.

\section{Discontinuities}
\label{sec:Disc-and-imaginary-for-Wilsonline}

As for their momentum-space analogs, Eq.\ (\ref{eq:lcopt-mtm-basic}), discontinuities 
in coordinate amplitudes given by
Eq.\ (\ref{eq:result-massless}) can be treated in some generality, both for Green functions and Wilson line expectations.   We can use the very general considerations above to give insight into the results of Refs.\ \cite{Laenen:2014jga,Laenen:2015jia}, which studied specifically how imaginary parts arise in the vacuum expectation values of products of Wilson lines evaluated in coordinate space.   This study resulted in a prescription for the imaginary parts of ladder diagrams that may be summarized as setting odd numbers of lines on the light cone.
We will see how these prescriptions arise in the formalism we have developed above, treating the very simplest case of massless  exchange for two Wilson lines joined at a singlet cusp, and also 
 a diagram with an internal vertex connected to three massless Wilson lines.   

\subsection{The imaginary part in four dimensions}

 Imaginary parts, or more specifically discontinuities, can arise in coordinate-space amplitudes only from the vanishing of denominators in Eq.\ (\ref{eq:result-massless}), through a cancellation between differences in external vertex minus positions and the corresponding path light-cone distance, $D^{(\pi_i)}_{(ba)}$, defined in Eq.\ (\ref{eq:deficit-def}).   Each light-cone distance $D^{(\pi_i)}_{(ba)}$  has a extremal value, which is determined by the four-vectors $x_b^\mu$ and $x_a^\mu$, and by the ordering of the $y_i^+$ along each path,
\bea
x_b^+ \ge y^+_{n_{ab}-1} \ge\ \cdots\ \ge y_2^+\ge x_a^+\, .
\eea
It is relatively easy to use the ordering of the $y_i^+$ to show that for given values of $x_a$ and $x_b$, the extremum of $D^{(\pi_i)}_{(ba)}$ is given by 
\bea
y_i^\mu -x_a^\mu\ =\ \alpha_i\, (x_b\ -\ x_a)^\mu\,  ,\ \mu=\perp\, ,\ +\, ,
\label{eq:y-prop}
\eea
for all $0<\alpha_i<\alpha_{i+1}<1$, that is, for all the $y_i^\mu$ placed along the direction between $x_a$ and $x_b$.  These configurations actually correspond to pinch singularities of coordinate-space amplitudes when $(x_a-x_b)^2=0$ and a set of internal vertices are lightlike separated. Nonetheless, no singularity is present even if internal vertices align along the direction between the external vertices unless the latter are relatively on the light cone~\cite{Erdogan:2013bga,Erdogan:2014gha}. 

From these considerations, we can study the denominators of (\ref{eq:result-massless}), denoted as
\bea
\Delta_{(ba)}^{(\pi_I)}\ \equiv
x^-_b\ -\ x^-_a\ -\ D^{(\pi_I)}_{(ba)}\ =\ \frac{\left( x_{b\perp}-x_{a\perp}\right)^2\ +\ \left(x_b-x_a\right)^2}{2\left(x_b^+-x_a^+\right)}\ -\ D^{(\pi_I)}_{(ba)}\, .
\label{eq:denominator-expand}
\eea
For spacelike separations $\left(x_b-x_a\right)^2<0$, this quantity is negative when the identities of (\ref{eq:y-prop}) are satisfied, and remains negative for all other values of $y_i^+$ and $y_{i\perp}$.   We encounter imaginary parts, then, only when external vertices are lightlike- or timelike-separated.   The $\Delta$'s, can be thought of as ``light-cone distance deficits", analogous to the ``light-cone energy deficits" of light-cone ordered perturbation theory in momentum space.

As noted above, there are a number of equivalent prescriptions for computing the imaginary parts of the individual terms $G_\P^{(\pi_I)}$.   We order the paths $P^{(\pi_I)}_{(ba)}$ in Eq.\ (\ref{eq:result-massless}) arbitrarily, denoting them as $P_A$, $A=1,\dots ,N_G$, where $N_G=L-N=m+n -1+K$, and where as above, $L$, $N$ and $K$ are the number of lines, internal vertices and loops in the ordered diagram $G$, which has  $m$ initial external vertices and $n$ final external vertices.     In this notation, Eq.\ (\ref{eq:result-massless}) becomes 
  \bea
G_\P^{(\pi_I)}\left (\{x_d\}_{\rm out},\{x_c\}_{\rm in}\right )\ & = & (2\pi )^N\ \frac{(-g)^N}{(4\pi^2)^L}\, \int\left(\prod_{i\in N} d^3y_i\right) \prod_{{\rm all}\ j} \frac{\theta(z_j^+)}{2z_j^+}\  
 \prod_{\{P_{A}^{(\pi_I)}\}\ A=1}^{N_G}\   \frac{-1}{\Delta_A^{(\pi_I)} -i\ep }\, ,
 \label{eq:result-in-Delta}
\eea 
In analogy to the discussion in momentum space, we define ``path-cut" coordinate diagrams for each covering set by
\bea
G_{\P,B}^{(\pi_I)}\left (\{x_d\}_{\rm out},\{x_c\}_{\rm in}\right )\ &=&\ (2\pi )^N\ \frac{(-g)^N}{(4\pi^2)^L}\, \int\left(\prod_{i\in N} d^3y_i\right) \prod_{{\rm all}\ j} \frac{\theta(z_j^+)}{2z_j^+}\  \nn\\
&\ &\hspace{10mm} \times\ 
 \prod_{\{P_{C}^{(\pi_I)}\}\ C=B+1}^{N_G}\   \frac{-1}{\Delta_C^{(\pi_I)} + i\ep }\ \left( -\ 2\pi\,  \delta\left( \Delta_B\right) \right) \  \prod_{\{P_{A}^{(\pi_I)}\}\ A=1}^{B-1}\   \frac{-1}{\Delta_A^{(\pi_I)} -i\ep }\, ,
\label{eq:cut-G}
\eea
where one light-cone path deficit is set to zero by the delta function, and other paths in the ordered set are assigned positive or negative imaginary parts.
Unlike momentum space, however, the ``light-cone paths" do not separate the diagram into disjoint parts in general, unless the diagram is planar.   Nevertheless, as in momentum space, using the identity $1/(x-i\ep)-1/(x+i\ep)=2\pi i\delta(x)$ we find an identity relating the imaginary part of the diagram to the sum over its path cuts for a given ordering,
\bea
2\ { \rm Im} G_{\P}^{(\pi_I)}\left (\{x_d\}_{\rm out},\{x_c\}_{\rm in}\right )\ &=& \ \sum_{B=1}^{N_G} G_{\P,B}^{(\pi_I)}\left (\{x_d\}_{\rm out},\{x_c\}_{\rm in}\right )\, .
\label{eq:Im-sum-cuts}
\eea
Notice that this relation holds for any choice of ordering for the paths.   This may seem surprising at first, since in momentum space time ordering appears to determine the ordering implemented in the unitarity identity in Eq.\ (\ref{eq:momentum-unitarity}) uniquely.   Nevertheless, permutations of momentum-space denominators in the cut diagrams, Eq.\ (\ref{eq:momentum-cut}) leave the form of the identity unchanged, at least for scalar theories, where we do not need to specify how the permutations act on Dirac products or other numerator factors.   In fact, this should be expected, because the unitarity will hold for any hermitian interaction Lagrangian, which can in principle directly mix states of arbitrary particle content.   

Equations (\ref{eq:result-in-Delta})--(\ref{eq:Im-sum-cuts}) are quite general and apply to any diagram.  An alternative expression for the imaginary part of Wilson line diagrams has been identified in Refs.\ \cite{Laenen:2014jga,Laenen:2015jia}, and illustrated in a large class of diagrams.   We now turn to a comparison of the two prescriptions, which are in fact consistent.

\subsection{Discontinuities in Wilson lines:  ladders}

While on a diagram-by-diagram basis, the equivalence of the light-cone ordered with the invariant form is clear for ladder diagrams in Eq.\ (\ref{eq:ladder-equivalence}), it remains for us to verify that the expression derived above for the imaginary part of an arbitrary diagram is consistent with the rules found in Refs.\ \cite{Laenen:2014jga,Laenen:2015jia}.  This is indeed the case, as may be seen by identifying light-cone ordered and covariant denominators for this class of diagrams, and by combining Eqs.\ (\ref{eq:cut-G}) and (\ref{eq:Im-sum-cuts}) to get
\bea
i\ { \rm Im} G_{\P,B}^{(\pi_I)}\left (\{x_d\}_{\rm out},\{x_c\}_{\rm in}\right )\ &=& \ (2\pi )^N\ \frac{(-g)^N}{(4\pi^2)^L}\, \sum_{B=1}^{N_G} \int\left(\prod_{i\in N} d^3y_i\right) \prod_{{\rm all}\ j} \frac{\theta(z_j^+)}{2z_j^+}\  \nn\\
&\ &\hspace{10mm} \times\ 
 \prod_{\{P_{C}^{(\pi_I)}\}\ C=B+1}^{N_G}\   \frac{-1}{\Delta_C^{(\pi_I)} + i\ep }\ \left( -i\pi\, \delta \left( \Delta_B\right) \right) \  \prod_{\{P_{A}^{(\pi_I)}\}\ A=1}^{B-1}\   \frac{-1}{\Delta_A^{(\pi_I)} -i\ep } \ . 
\label{eq:Im-sum-cuts-explicit}
\eea
Following Ref.\ \cite{Laenen:2014jga}, we expand each denominator into a principle value plus a delta function
\bea
\frac{1}{\Delta_E^{(\pi_I)} \pm i\ep }\ =\ {\rm PV}\ \frac{1}{\Delta_E^{(\pi_I)}}\ \mp\ i\pi \delta\left( \Delta_E \right)\, . \label{eq:PVdelta}
\eea
The imaginary part in Eq.\  (\ref{eq:Im-sum-cuts-explicit}) is given by the sum over all terms with products of odd numbers of delta functions.   Suppose we identify any set $D_M$ of $M$ such terms, corresponding to path denominators $\{ \Delta_k\}$, $k=1, \dots , M$, including the explicit delta function $\delta(\Delta_B)$, with the index following the (arbitrary) ordering in Eq.\ (\ref{eq:Im-sum-cuts-explicit}).  Because the remaining denominators are all principle values, the contribution of $D_M$ is imaginary when $M$ is an odd number.   For even $M$, the contribution should vanish.   We can write the full sum in (\ref{eq:Im-sum-cuts-explicit}) as a sum over sets $D_M$, and for each such set a sum over terms in which definite numbers of the ordered delta functions appear with $\pm i\pi$, arising from $\mp i\ep$ in the corresponding denominator.    The result of this expansion can be written as
\bea
i\ { \rm Im} G_{\P}^{(\pi_I)}\left (\{x_d\}_{\rm out},\{x_c\}_{\rm in}\right )\ &=& \ 
(2\pi )^N\ \frac{(-g)^N}{(4\pi^2)^L}\, \sum_{B=1}^{N_G} \int\left(\prod_{i\in N} d^3y_i\right) \prod_{{\rm all}\ j} \frac{\theta(z_j^+)}{2z_j^+}\  \nn\\
&\ &\hspace{-10mm} \times\ 
\sum_{M=1}^{N_G}
 \prod_{C \in \{P^{(\pi_I)}\} / D_M }\ {\rm PV} \frac{-1}{\ \ \Delta_C^{(\pi_I)}\ }  \prod_{i\in D_M =1}^M  \ \delta\left( \Delta_i\right) \  \times\
 (i\pi)^M\ \left [ \sum_{i=1}^{M} (-1)^i \right] \ . 
\label{eq:Im-sum-cuts-compare}
\eea
The final factor equals $-1$ for $M$ odd, and zero for $M$ even.
For ladder exchanges between Wilson lines, this result for the imaginary part is equivalent to the prescription of Ref.\ \cite{Laenen:2014jga,Laenen:2015jia}, in which the imaginary part is the sum of all terms with an odd number of exchanged gluons with propagators replaced by $i\pi$ times a delta function.

\subsection{Discontinuities in Wilson lines:  the three-gluon vertex}

\begin{figure}

\centering
\includegraphics[height=4cm]{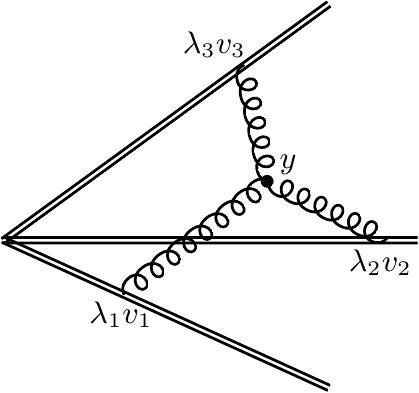}

\caption{The diagram with an internal three-gluon vertex joining three Wilson lines. \label{fig:3g-vertex}}

\end{figure}

We now proceed with a representative example of gluon exchange with interactions, also discussed in Refs.\ \cite{Laenen:2014jga,Laenen:2015jia}, where an internal three-gluon vertex connects three Wilson lines in different directions~$v_i$ as shown in Fig.~\ref{fig:3g-vertex}. We will compute this diagram in coordinate space following our prescription. 
We choose $\lambda_i v^{\mu}_i$ as a position along the $i$th line and $y^\mu$ as the position of the internal vertex, and then write the diagram as an integral over $y^\mu$ and the $\lambda_i$s,
\bea
F^{(3)}(v_i^\mu,\vep)\ =\  \int d^Dy\ \prod_{i=1}^3 \int_0^\infty d\lambda_i\ V \left(v_i,\partial_y \right) \ , 
\eea
where the integrand $V(v_i,\partial_y)$ involves the directions $v_i$ and the derivative operator at the internal vertex that acts on the three propagators. Suppressing overall color and other constant factors, the integrand can be written as \cite{Mitov:2009sv} 
\be 
V(v_i,\partial_y) = -ig^4\sum^3_{i,j,k=1} \vep_{ijk}\, v_i\cdot v_j\,\frac{1}{-(\lambda_iv_i-y)^2+ i\ep} \frac{1}{-(\lambda_jv_j-y)^2+i\ep}  \,v_k\cdot \partial_y \frac{1}{-(\lambda_kv_k-y)^2+i\ep} \ . 
\ee
We fix the points along the Wilson lines and choose a specific ordering of the $x_i^+\equiv \lambda_i v_i^+$ and $y^+$.  Using the general result of Eq.\ (\ref{eq:hatw-deriv}), we write the three-gluon vertex as a differential operator acting on the $y^-$~integral of 
the scalar 3-point diagram, just as we did for the scalar vertex function in Eq.~(\ref{eq:scalar-vertex-hatz}),
\bea 
F^{(3)}(v_i^\mu,\vep)\ =\  \int d^3y\ \prod_i \int_0^\infty d\lambda_i\, {\cal V}\left(z^+_i,\hat z^-_i,\partial_{\hat z_i}\right)  \,  \gamma^{(3)}_{\rm scalar}(z_i) \ , 
\eea
with $z_i=x_i - y$. The operator ${\cal V}$, which involves derivatives and factors of coordinates, does not change the reality properties of the scalar function.
Here as above, the $y^-$ integral is absorbed into a scalar function, which, using the second form on the right-hand side of Eq.\ (\ref{eq:hatw-deriv}), we may express as
\bea
\gamma^{(3)}_{\rm scalar}(z_i)\ =\ \int dy^-\ \frac{1}{-(\lambda_iv_i-y)^2+i\ep}\ \frac{1}{-(\lambda_jv_j-y)^2+i\ep}\   \frac{1}{-(\lambda_kv_k-y)^2+ i\ep}\, ,
\label{eq:gamma-3}
\eea
and from which we have factored all overall constants.

The integral in Eq.\ (\ref{eq:gamma-3}) depends on the relative orderings among $y^+$ and the plus coordinates of the vertices along the Wilson lines, $\lambda_iv_i^+$.   To be specific, let's choose the region
\bea
R_{3y21}\ \equiv\ \lambda_3v_3^+\ >\ y^+ >\ \lambda_2v_2^+\ >\ \lambda_1v_1^+\, .
\eea
Other regions are treated in the same fashion.   At fixed values of $y_\perp$ and $y^+$, we can apply the rules developed above to find the light-cone-ordered form for the scalar function in this region of plus ordering.    In the terminology of Sec.~\ref{sec:path-denominators}, there are two paths in the covering set from the ``initial" vertices, at $\lambda_1v_1$ and $\lambda_2v_2$ to the final vertex at 
$\lambda_3v_3$, both passing through the internal vertex at $y$.   There are, therefore, two denominators given by
\bea
\gamma^{(3)}_{\rm scalar}[R_{3y21}]\ =\ \frac{-2\pi i}{\prod_{i=1}^3 2|\lambda_iv_i^+  - y^+|}\
\frac{1}{\lambda_3v_3^-\ -\ D_{32}\  -\ \lambda_2v_2^-\ - i\ep}\ \frac{1}{\lambda_3v_3^-\ -\ D_{31} -\ \lambda_1v_1^-\ - i\ep}\, ,
\label{eq:R3y21-lco}
\eea
where the $D_{ij}$ are ``light-cone distances" defined as in Eq.\ (\ref{eq:deficit-def}),
\bea 
D_{ij} & =& \frac{(\lambda_iv_{i\perp}-y_{\perp})^2}{2(\lambda_iv^+_i-y^+)}  + \frac{(y_{\perp}-\lambda_jv_{j\perp})^2}{2(y^+ - \lambda_jv^+_j)}  \ . 
\label{eq:deeij}
\eea
This is particularly simple, and can also be understood in terms of a direct evaluation of the $y^-$ integral.  The ordering of plus components specifies that the $y^-$ pole in the $(\lambda_3v_3-y)^2$ denominator in the integrand of Eq.\ (\ref{eq:gamma-3}) is in the lower half plane, while the poles from the other two propagators are in the upper half plane.  Closing the $y^-$ contour in the lower half plane gives the result in Eq.\ (\ref{eq:R3y21-lco}) immediately.   Closing in the upper half plane gives an alternate expression with two terms, which, of course, is equal to the first expression.   Other orderings of plus components of the vertices will give alternate, but equally simple, expressions.

To compare with the results of Ref.\ \cite{Laenen:2014jga}, we can identify the real part of the product of denominators in (\ref{eq:R3y21-lco}) by expanding each of them into a real principal value and imaginary delta function.   The imaginary part can be found from replacing either of the two remaining denominators with a delta function, which would correspond to vanishing light-cone deficits between the light-cone distances computed along a (two-gluon) path between two of the ``external" vertices along the Wilson lines.   We must observe, however, that  the principal value prescription is completely well defined only if the $\lambda$ integrals are done first, before the integrals over the internal vertices, just as presented in Ref.~\cite{Laenen:2014jga}.   As a practical matter in our case, evaluating the $y^-$ integral at one of the denominator poles, as we do here, can shift the sign of the imaginary part in the other denominators. Applying a principal value prescription in (\ref{eq:R3y21-lco}) can be expected to agree term by term with the results of \cite{Laenen:2014jga} only when its remaining denominators are finite.   When carried out consistently, of course, our method and that of Ref.\ \cite{Laenen:2014jga} must give the same complete answer.

In fact, we readily verify that if we identify the denominators of the light-cone result  (\ref{eq:R3y21-lco}) with principal values of propagators, the result is equivalent to the contribution to the imaginary part of Fig.\ \ref{fig:3g-vertex} found in Ref.\ \cite{Laenen:2014jga}, that comes from adding the three single-gluon poles for this range of plus coordinates.   For us, however, away from the poles in the remaining denominators, this is a contribution to the relatively real part of the diagram.  This difference is simply due to the overall normalization.  Equation (\ref{eq:R3y21-lco}), in fact has both real and imaginary parts.   Independently of the normalization, however, the discontinuity of the diagram in Fig.\ \ref{fig:3g-vertex} is found by applying Eq.\ (\ref{eq:Im-sum-cuts}) to Eq.\ (\ref{eq:R3y21-lco}), 
\bea
2i\,{\rm Disc}\ \gamma^{(3)}_{\rm scalar}  &=& 
  \frac{-2\pi i  }{\prod_{i=1}^3 2|\lambda_iv_i^+  - y^+|}\left[ 
\frac{1}{\lambda_3v_3^-  -\ \lambda_2v_2^- -D_{32} - i\ep}\ \frac{1}{\lambda_3v_3^- - \lambda_1v_1^- - D_{31} - i\ep} \right. \nn \\
& \ & \hspace{3cm} \left. - 
\frac{1}{\lambda_3v_3^-  - \lambda_2v_2^- -D_{32} + i\ep}\ \frac{1}{\lambda_3v_3^- - \lambda_1v_1^- - D_{31}  + i\ep}\right] \nn \\
& =&   \frac{2\pi^2 }{\prod_{i=1}^3 2|\lambda_iv_i^+  - y^+|}\left[ \ 
\frac{1}{\lambda_3v_3^-  -\lambda_2v_2^- - D_{32} + i\ep}\ \delta\left(\lambda_3v_3^-  - \lambda_1v_1^- - D_{31} \right) \right. \nn \\
& \ & \hspace{3cm} \left. + \ 
\delta\left(\lambda_3v_3^-    -  \lambda_2v_2^- - D_{32} \right)\ \frac{1}{\lambda_3v_3^- -  \lambda_1v_1^- - D_{31}-i\ep } \right]  \ . 
\label{eq:R3y21-cut}
\eea
For Fig.\ \ref{fig:3g-vertex}, the discontinuity sets two lines to the light cone, corresponding to vanishing light-cone deficits for a path between the diagram's external vertices.   
For the diagram with an internal vertex, as in diagrams with ladder exchange, physical discontinuities are thus produced from coordinate configurations in which a full path of gluon lines on the light cone connects external vertices.  For the ordering we have shown here, there are two such paths, corresponding to the two terms in (\ref{eq:R3y21-cut}).   An additional contribution, real here and imaginary in \cite{Laenen:2014jga}, comes when both denominators are replaced by delta functions in Eq.\ (\ref{eq:R3y21-lco}).

\section{Massive Lines and Dimensional Continuation}
\label{sec:mass-dim}

So far, our discussion has treated massless scalar theories in four dimensions.   We have used the four-dimensional form of the coordinate-space propagator in an essential way, in particular, that it has two simple poles on the light cone, and no branch cuts.   This enables us to derive  Eq.\ (\ref{eq:E-integrals-1}) from Eq.\ (\ref{eq:delta-to-E}), in terms of  residues of poles in the $z^-$ coordinates for each line.   The analysis must clearly change for massive fields, for which the propagators are given in terms of Bessel functions even in four dimensions, and also for massless fields in $4-2\vep$ dimensions, where the propagator is
\bea
\Delta(z^2,m=0)\ =\  \frac{\Gamma(1-\vep)}{4\pi^{2-\vep}}\ \frac{1}{\left(-z^2\ +\ i\ep\right)^{1-\vep}}\, ,
\eea
which has a branch cut from $0\le z^2<\infty$.  These essential complications in coordinate space are to be contrasted with the generality of light-cone perturbation theory results in momentum space, where masses and dimensional continuation can be implemented directly in the integrand, and in the definition of transverse integrations, respectively.      Here we briefly sketch the necessary generalization for coordinate space.

This generalization requires the introduction of extra dispersive integrals that replace the simple poles above, while respecting the basic structure of the light-cone results.  Both for masses, $m$ and dimensions $\vep\ne 0$, the analog of Eq.\ (\ref{eq:G-massless}) for Green functions in coordinate representation, but now with scalar lines of mass $m$, can be constructed from propagators written in dispersive form, 
\bea
\Delta(z^2,m)\ =\ \int_0^\infty \frac{dz'{}^2}{\pi }\; \frac{ {\rm Im}\; \Delta \left(z'{}^2+i\ep,m \right)}{-2z^+z^- + z_\perp^2 + z'{}^2+i\ep}\, .
\eea
To be specific, the imaginary parts for the massless and massive scalar propagators in $4-2\vep$ dimensions are given by \cite{Zhang:2008jy}
\bea
{\rm Im} \Delta(z^2,m)\ &=& \theta\left(z^2\right)\  \frac{ m^{1-\vep} }{2^{3-\vep}\pi^{1-\vep} \left(\sqrt{z^2}\right)^{1-\vep}}\ H^{(2)}_{1-\vep} \left (m\sqrt{z^2} \right)
\nonumber\\
{\rm Im} \Delta(z^2,0)\ &=& \ \frac{1}{(z^2)^{1-\vep}}\ \frac{\pi}{\Gamma(\vep)}\, ,
\eea
with $H_{1-\vep}$ a Bessel function.

In place of the generic form in Eq.\ (\ref{eq:G-massless}) for a Green function with massless lines, we now have
\bea
G_E(\{ x_a\}) \ &=& \ \frac{(-ig)^N}{(4\pi^2)^L}\;
\prod_{\mathrm{lines}\, j=1}^L \int_0^\infty  \frac{dz_j'{}^2}{\pi }\; 
\ {\rm Im}\; \Delta \left(z_j'{}^2+i\ep \right)
\nn\\
&\ & \hspace{10mm} \times\ 
\int\prod_{\mathrm{vertices}\, i=1}^N d^4y_i \prod_{\mathrm{lines}\, j=1}^L 
\frac{1}{-z^2_j(y_i,x_a)\ +\ z'_j{}^2 + i\ep} \ .
\label{eq:G-massive} 
\eea
In this form, the full reasoning leading to Eq.\ (\ref{eq:result-massless}) can be applied to the integrals over the $y_i$'s, leading to the same result for fixed values of the variables $z_j'{}^2$.   The effect of the dispersion form for each propagator is to change the function $D$ by the addition of the variable $z_j'{}^2$ to the dependence on transverse coordinates,
\bea
\bar D^{(\pi_I)}_{(ba)}\ =\ \sum_{i\in P_{(ba)}^{(\pi_I)}} \frac{(y_{i\perp}\ \ -\ y_{i-1\perp})^2+ z_{(i,i-1)\perp}^{'2}}{2(y^+_i-y^+_{i-1})}\, .
\eea
Here $z_{(i,i-1)\perp}^{'2}$ is the dispersive variable corresponding to the line $z_{(i,i-1)}=y_i-y_{i-1}$ along this path.   As the mass vanishes, and as the number of dimensions approaches four, the discontinuity of $\Delta$ vanishes away from the light cone  and approaches a delta function, reproducing the massless result.   

\section{Conclusions}

We have developed a coordinate-space analog of the venerable light-cone-ordered perturbation theory in momentum space, finding that the central role of states in momentum space is taken in coordinate space by paths between external operators.  Discontinuities in amplitudes can arise when these points are separated by lightlike distances, corresponding to discontinuities in momentum space associated with states that satisfy 
energy conservation.    This principle sheds light on the discontinuities of expectation values of Wilson lines, which can be evaluated in momentum or coordinate space.    We anticipate applications as well to the interpretation of infrared singularities of cross sections and their cancellation, which are also accessible in a coordinate formulation \cite{Erdogan:2016ylj}.

The detailed prescriptions developed above require the identification of ``covering sets" of paths between ``incoming" and ``outgoing" vertices, for each partially ordered set of internal and external vertices.   The possibility of a geometrical generalization of these results applied to high orders is intriguing, and we hope to take it up in future work.

\acknowledgments
This work was supported in part by the National Science Foundation,  grants 
PHY-1316617 and 1620628. The work of O.E. was also supported by  the Department of Energy  under \mbox{DE-SC$0010118$}.  We thank Stanley J. Brodsky, Eric Laenen, and Kasper Larsen for very useful conversations.

\begin{appendix}

\section{Paths from direct evaluation}

In this appendix, we provide an alternative proof of Eq.\ (\ref{eq:result-massless}) for all massless lines, by the direct evaluation of $y^-$ integrals.  As in Sec.\ \ref{subsec:construction}, this argument is iterative in nature, this time in the order of the diagram, for arbitrary numbers of initial and final vertices.   It applies equally to tree and loop diagrams.   

At zeroth order in the interaction, an arbitrary diagram is simply a set of $n$ lines, each connecting an initial and a final vertex. For fixed values of the $x_i^+$ of these  $2n$ vertices, Eq.\ (\ref{eq:result-massless}) applies immediately with no integrations over internal vertices.   We then assume the result for all diagrams of some fixed order $N>0$.   We  consider an arbitrary ordered diagram of order $N+1$, for any ordering of its $N+1$ internal vertices, which we label as $y_{N+1}^+ > y_N^+ > \dots > y_1^+$.   In this diagram, the vertex at $y_{N+1}$ is connected to $d_{N+1}$ predecessor lines (which can arise from internal or initial predecessor vertices) and is a predecessor of another set $p_{N+1}$ of descendant vertices, which by construction are final vertices.   In this ordering, $d_{N+1}+p_{N+1}$ is the number of lines that are attached to the vertex $y_{N+1}$.   
 We refer to this diagram as ${\cal G}^{(\pi_I,N+1)}_{(n,m)\P} \left(\{X_1, \dots , X_n\}_{\rm out},\{z_1 ,\dots , z_m\}_{\rm in}\right)$.   

\begin{figure}

\centering
\includegraphics[height=2.4cm]{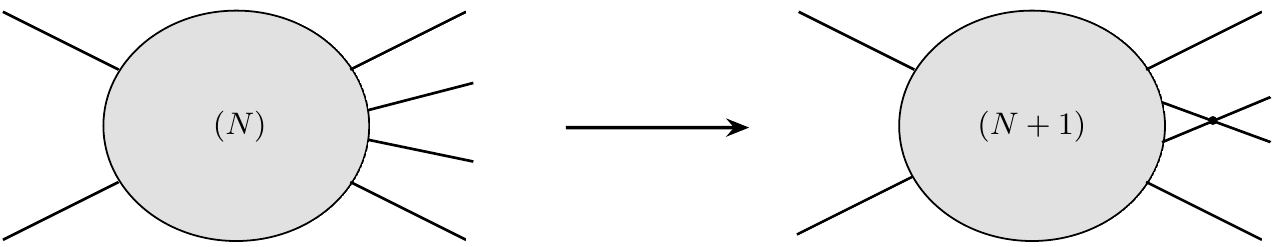}

\caption{Diagram representing the addition of an interaction vertex. \label{fig:iterative}}

\end{figure}

We can express our $N+1$st order diagram, ${\cal G}^{(\pi_I,N+1)}_{(n,m)\P}$ in terms of an $N$th order diagram, ${\cal G}^{(\pi_I,N)}_{(n-p_{N+1}+d_{N+1},m)\P} $ by identifying $d_{N+1}$ final vertices of the $N$th order diagram at point $y_{N+1}$, and multiplying by $p_{N+1}$ propagators that connect the new vertex at $y_{N+1}$ with new external vertices at points $X_{n-p_{N+1}+1}, \dots, X_n$, all with $X_j^+>y_{N+1}^+$.   Schematically, the relation can be written as
\bea
{\cal G}^{(\pi_I,N+1)}_{(n,m)\P} \left(\{X_1, \dots , X_{n-p_{N+1}+1}, \dots , X_n \}_{\rm out}, \{z_1 , \dots ,  z_m \}_{\rm in}\right)
 &=& 
\int d^4y_{N+1}\
\prod_{j=n-p_{N+1}+1}^n\ \frac{1}{ -(X_j-y_{N+1}  )^2+ i\ep}
\nn\\
&\ & \hspace{-70mm} \times\
{\cal G}^{(\pi_I,N)}_{(n-p_{N+1}+d_{N+1},m)\P} \left(\{X_1 ,  \dots ,  y_{N+1} ,\dots , y_{N+1} , \dots ,  X_{n-p_{N+1}+d_{N+1}}\}_{\rm out},\{z_1,  \dots ,  z_m\}_{\rm in}\right)\, ,
\label{eq:iterative-yminus}
\eea
where $d_{N+1}$ of the external vertices are set to $y_{N+1}$. As usual, the subscript ${\cal P}$ indicates that the plus coordinates of each diagram are (partially) ordered, with the partial ordering of ${\cal G}^{(\pi_I,N+1)}_{(n,m)\P}$ being determined by the partial ordering of ${\cal G}^{(\pi_I,N)}_{(n-p_{N+1}+d_{N+1},m)\P}$, along with the relative orderings of $y_{N+1}^+$ and of its descendant vertices, which are all final vertices. 
An example of the relationship in Eq.\ (\ref{eq:iterative-yminus}) is shown in Fig.\ \ref{fig:iterative}, where $d_{N+1}=p_{N+1}=2$.

Our aim is to do the $y_{N+1}^-$ integral in Eq.\ (\ref{eq:iterative-yminus}), and to show that if ${\cal G}^{(\pi_I,N)}_{(n-p_{N+1}+d_{N+1},m)\P} $ is of the form of Eq.\ (\ref{eq:result-massless}), with paths that extend from its initial to its final vertices, so is ${\cal G}^{(\pi_I,N+1)}_{(n,m)\P}$.   This result will follow by keeping track of the $y_{N+1}^-$ poles  in this expression, because all $y_{N+1}^-$-dependence is either in the explicit propagators in (\ref{eq:iterative-yminus}), or in the denominators of the $N$th order diagram ${\cal G}^{(\pi_I,N)}_{(n-p_{N+1}+d_{N+1},m)\P}$, 
which obeys (\ref{eq:result-massless}).

By our iterative assumption, in function ${\cal G}^{(\pi_I,N)}_{(n-p_{N+1}+d_{N+1},m)\P}$, $y^-$ appears in denominators that are specified by paths from the initial vertices of the diagrams, to the set of $d_{N+1}$ final state vertices that have been fused into the new interaction vertex at $y_{N+1}^\mu$.   We denote the number of $y^-_{N+1}$-dependent denominators  by $m_G$.   We always have $m_G\ge d_{N+1}$. The other denominators of ${\cal G}^{(\pi_I,N)}_{(n-p_{N+1}+d_{N+1},m)\P}$ are independent of $y^-$ and do not enter into the argument.  

The general form of $y_{N+1}^-$-dependent denominators in function ${\cal G}^{(\pi_I,N)}_{(n-p_{N+1}+d_{N+1},m)\P}$ is given by Eq.\ (\ref{eq:result-massless}) and (\ref{eq:deficit-def}) as
\bea
 y_{N+1}^-\ -\ \ z_a^-\ -\  \sum_{{\rm vertices}\ i\in P_{(y_{N+1}a)}} \frac{(\ y_{i\perp}\ \ -\ y_{i-1\perp})^2}{2(y^+_i-y^+_{i-1})}\ -\ i\ep
 \ \equiv\
 y_{N+1}^-\ -\ \delta_a -\ i\ep\, ,
\eea
where  the right-hand side serves as a definition of $\delta_a$, and where $\{P_{(y_{N+1}a)}\}$ is the relevant set of paths from initial vertices at points $z_a$ to the final vertex at point $y_{N+1}$.   The poles of all these denominators are in the upper $y_{N+1}^-\equiv y$ half-plane.   Their number, and the real values of the locations of their poles depend on the details of the diagram, but we shall only need to assume that there is a finite number of them, which we denote $m_G$.   The explicit poles of the new propagators in (\ref{eq:iterative-yminus}) at points we label $\xi_l$, in contrast, are all in the $y_{N+1}^-$ lower half plane, because by construction every $X_j^+>y_{N+1}^+$.   Suppressing all overall factors, the general $y_{N+1}^-$ integral is therefore of the form
\bea
I_{p_{N+1},m_G}\ =\ \frac{1}{2\pi i}\ \int_{-\infty}^\infty d y \ \prod_{l=1}^{p_{N+1}} \frac{1}{\xi_l\ -\ y\ -\ i\ep}\ \prod_{i=1}^{m_G} \frac{1}{y\ -\ \delta_i\ -\ i\ep}\, ,
\label{eq:I-pN-mG}
\eea
with the normalization chosen for convenience.
In the following, we further simplify the notation by suppressing the infinitesimal imaginary parts of the denominators, which are all negative.

We confirm below that  the result of the $y_{N+1}^-$ integral can be written as a product of denominators of the same form, which depend on paths leading from the same set of initial vertices $\{z_a\}$ to the ``new" $p_{N+1}$ vertices $X^\mu_j$ connected to $y_{N+1}^-$ by the explicit propagators in Eq.\ (\ref{eq:iterative-yminus}).   For renormalizable theories we would need to show that this holds for $p_{N+1}$ from one to three, but in fact it holds for any value of $p_{N+1}$.   

The result is immediate for $p_{N+1}=1$, because when we close in the lower half plane, the value of $y_{N+1}^-$ at the pole adds precisely the extra ``length" to the path from each initial vertex $z_a$ to $y_{N+1}$ to produce the denominator for the path from $z_a$ to the single new vertex $X_n$.   The cases starting $p_{N+1}=2$ require algebraic manipulation.

\subsection{$m_G$ to 2 Identity}

The $y$ integral when there are two descendant vertices for the additional interaction is found from Eq.\ (\ref{eq:I-pN-mG}) with $p_{N+1}=2$,  
\bea
I_{2,m_G}\ \equiv\ \frac{1}{2\pi i}\, \int_{-\infty}^\infty dy\ \left( \frac{1}{\xi_2-y}\, \frac{1}{\xi_1-y} \right)\
\prod_{i=1}^{m_G}\ \frac{1}{y-\delta_i}\, .
\eea
Closing the $y$ contour in the lower half-plane for this integral (recalling that the denominators all have negative imaginary parts), we pick up the poles of the propagators connecting $y_{N+1}$ and its descendant vertices to find,
\bea
I_{2,m_G}\ =\ 
\frac{1}{\xi_1-\xi_2}\ \left( \prod_{i=1}^{m_G}\ \frac{1}{\xi_2 -\delta_i}\ -\ \prod_{i=1}^{m_G}\ \frac{1}{\xi_1 - \delta_i} \right)\, .
\label{eq:m-to-2-yint}
\eea
This, of course, is not yet the desired result.   To derive an expression with only forward-moving paths, we add zero to this expression, in the form,
\bea
0\ =\ \frac{1}{\xi_1-\xi_2}\ \left( -\ \sum_{l=1}^{m_G-1}\ \prod_{j=l+1}^{m_G}\ \frac{1}{\xi_2-\delta_j}\ \prod_{i=1}^{l}\ \frac{1}{\xi_1-\delta_i}
\  +\
\sum_{l=1}^{m_G-1}\ \prod_{j=l+1}^{m_G}\ \frac{1}{\xi_2-\delta_j}\ \prod_{i=1}^{l}\ \frac{1}{\xi_1-\delta_i}  \right)\, .
\label{eq:add-zero}
\eea
From these sums, we isolate the $l=1$ term from the first sum, and $l=m_G-1$ from the second, and group them with the $\xi_2$ product and the $\xi_1$ product in Eq.\ (\ref{eq:m-to-2-yint}) respectively.   This gives
\bea
I_{2,m_G}\ &=&\
\frac{1}{\xi_1-\xi_2}\ \left(  \prod_{j=1}^{m_G}\ \frac{1}{\xi_2 -\delta_j}\  -\   \frac{1}{\xi_1-\delta_1}\  \prod_{j=2}^{m_G}\ \frac{1}{\xi_2-\delta_j} \right)  
\nn\\
&\ & \hspace{5mm} 
+\ \frac{1}{\xi_1-\xi_2}\ \left( \frac{1}{\xi_2-\delta_{m_G}} \ \prod_{i=1}^{m_G-1}\ \frac{1}{\xi_1 - \delta_i}  \ -\ \prod_{i=1}^{m_G}\ \frac{1}{\xi_1 - \delta_i} \right)
\nn\\
&\ & \hspace{5mm} +\ 
\frac{1}{\xi_1-\xi_2}\ \left( -\ \sum_{l=2}^{m_G-1}\ \prod_{j=l+1}^{m_G}\ \frac{1}{\xi_2-\delta_j}\ \prod_{i=1}^{l}\ \frac{1}{\xi_1-\delta_i}
\  +\
\sum_{l=1}^{m_G-2}\ \prod_{j=l+1}^{m_G}\ \frac{1}{\xi_2-\delta_j}\ \prod_{i=1}^{l}\ \frac{1}{\xi_1-\delta_i}  \right)\, .
\eea
We next combine the first pairs of terms, which cancels the $\xi_1-\xi_2$ denominators in these expressions.  Also, we relabel the indices of the first summation (by $l\rightarrow l+1$).  Then, both sums run from $l=1$ to $m_G-2$, so that they can be grouped into a single summation,
\bea
I_{2,m_G}\ &=& \ \prod_{j=1}^{m_G}\ \frac{1}{\xi_2 -\delta_j}\ \frac{1}{\xi_1-\delta_1}\ +\ \frac{1}{\xi_2-\delta_{m_G}}\ \prod_{i=1}^{m_G}\ \frac{1}{\xi_1 -\delta_i}\
\nn\\
&\ & \hspace{-5mm} +\ 
\frac{1}{\xi_1-\xi_2}\ \sum_{l_1=1}^{m_G-2}\ \left( -\  \prod_{j=l+2}^{m_G}\ \frac{1}{\xi_2-\delta_j}\  \frac{1}{\xi_1-\delta_{l+1}}\ \prod_{i=1}^{l}\ \frac{1}{\xi_1-\delta_i}
\  +\
 \prod_{j=l+2}^{m_G}\ \frac{1}{\xi_2-\delta_j}\ \frac{1}{\xi_2-\delta_{l+1}}\ \prod_{i=1}^{l}\ \frac{1}{\xi_1-\delta_i}  \right)
 \nn\\
 &=& \ \prod_{j=1}^{m_G}\ \frac{1}{\xi_2 -\delta_j}\ \frac{1}{\xi_1-\delta_1}\ +\ \frac{1}{\xi_2-\delta_{m_G}}\ \prod_{i=1}^{m_G}\ \frac{1}{\xi_1 -\delta_i}\
 +\ 
 \sum_{l=1}^{m_G-2}\  \prod_{j=l+2}^{m_G}\ \frac{1}{\xi_2-\delta_j}\  \frac{1}{\xi_2-\delta_{l+1}}\ \frac{1}{\xi_1-\delta_{l+1}}\ \prod_{i=1}^{l}\ \frac{1}{\xi_1-\delta_i}  \, ,
 \nn\\
\eea
where in the second equality we combine the final two terms under the remaining summation.    This can be reorganized as
\bea
I_{2,m_G}\ =\  \sum_{l=1}^{m_G}\ \prod_{j=l}^{m_G} \frac{1}{\xi_2-\delta_j}\ \prod_{i=1}^l \frac{1}{\xi_1-\delta_i}\, .
\label{eq:I-2-result}
\eea
In this expression every denominator reflects a path from an initial vertex, through the vertex at $y_{N+1}$ to one of its descendant vertices, $X_1$ or $X_2$.   This is the result we were after.   We note that, as in the construction of Sec.\ \ref{subsec:construction}, the procedure leading from Eq.\ (\ref{eq:add-zero}) to (\ref{eq:I-2-result}) introduces an arbitrary ordering for the paths.

\subsection{$m_G$ to 3 identity}

In this case, we start with the result we wish to prove, which generalizes the $p_{N+1}=2$ relation, Eq. (\ref{eq:I-2-result}).   As in the previous case, this involves an arbitrary ordering of the paths that arrive at vertex $y_{N+1}$,  
\bea
I_{3,m_G}\ &\equiv&\ \sum_{m\ge l_2\ge l_1\ge 1}\ \prod_{k=l_2}^{m_G}\ \frac{1}{\xi_3-\delta_k}\ \prod_{j=l_1}^{l_2}\ \frac{1}{\xi_2-\delta_j}\ \prod_{i=1}^{l_1}\ \frac{1}{\xi_1-\delta_i}\, .
\label{eq:I-3-result}
\eea
Working backward, we recognize that the right pair of products is $I_{2,l_2}$, in the form of Eq.\ (\ref{eq:I-2-result}), and then apply the equivalent expression for this quantity, Eq.\ (\ref{eq:m-to-2-yint}),
\bea
I_{3,m_G}\  &=&\
\sum_{l_2= 1}^{m_G}\ \prod_{k=l_2}^{m_G}\ \frac{1}{\xi_3-\delta_k}\ \sum_{l_1=1}^{l_2}\ \prod_{j=l_1}^{l_2}\ \frac{1}{\xi_2-\delta_j}\ \prod_{i=1}^{l_1}\ \frac{1}{\xi_1-\delta_i}
\nn\\
&=& \ \sum_{l_2= 1}^{m_G}\ \prod_{k=l_2}^{m_G}\ \frac{1}{\xi_3-\delta_k}\ I_{2,l_2}
\nn\\
&=& \ 
\sum_{l_2= 1}^{m_G}\ \prod_{k=l_2}^{m_G}\ \frac{1}{\xi_3-\delta_k}\ \left\{ \frac{1}{\xi_1-\xi_2} \left[ \prod_{j=1}^{l_2} \frac{1}{\xi_2-\delta_j}\ -\
 \prod_{i=1}^{l_2} \frac{1}{\xi_1-\delta_i} \right] \right \}\, .
 \label{eq:I-3-iterative}
\eea
This  expression is a single sum over a combination of two products.
Factoring out the common $1/(\xi_1-\xi_2)$, we can again apply the $p_{N+1}=2$ identity that relates (\ref{eq:I-2-result}) to (\ref{eq:m-to-2-yint}),
\bea
I_{3,m_G}\ &=&\ \frac{1}{\xi_1-\xi_2}\ \left\{
\sum_{l_2=1}^{m_G} \prod_{k=l_2}^{m_G}\ \frac{1}{\xi_3-\delta_k}\ \prod_{j=1}^{l_2}\ \frac{1}{\xi_2-\delta_j}\
-\ \sum_{l_2=1}^{m_G} \prod_{k=l_2}^{m_G}\ \frac{1}{\xi_3-\delta_k}\ \prod_{i=1}^{l_2}\ \frac{1}{\xi_1-\delta_i}
\right\}
\nn\\
&=&\ \frac{1}{\xi_1-\xi_2}\ \left\{
\frac{1}{\xi_2-\xi_3}\ \left [  \prod_{k=1}^{m_G} \frac{1}{\xi_3-\delta_k}\ -\ \prod_{j=1}^{m_G}\ \frac{1}{\xi_2-\delta_j} \right ]\
-\ \frac{1}{\xi_1-\xi_3}\ \left [  \prod_{k=1}^{m_G} \frac{1}{\xi_3-\delta_k}\ -\ \prod_{i=1}^{m_G}\ \frac{1}{\xi_1-\delta_i} \right ]
\right \}
\nn\\
&=& \frac{1}{\xi_2-\xi_1}\, \frac{1}{\xi_3-\xi_1}\ \prod_{i=1}^{m_G}\ \frac{1}{\xi_1-\delta_i}\
+\ \frac{1}{\xi_1-\xi_2}\, \frac{1}{\xi_3-\xi_2}\ \prod_{j=1}^{m_G}\ \frac{1}{\xi_2-\delta_j}\ 
+\ \frac{1}{\xi_1-\xi_3}\, \frac{1}{\xi_2-\xi_3}\ \prod_{k=1}^{m_G}\ \frac{1}{\xi_3-\delta_k}\, .
\label{eq:I-3-3-terms}
\eea
where the three terms of the final expression correspond to the fourth, second, and the sum of the first and third terms of the foregoing expression, respectively.  We recognize this expression as the integral
\bea
I_{3,m_G}\ =\ \frac{1}{2\pi i}\, \int_{-\infty}^\infty dy\ \left( \frac{1}{\xi_3-y}\, \frac{1}{\xi_2-y}\, \frac{1}{\xi_1-y} \right)\
\prod_{i=1}^{m_G}\ \frac{1}{y-\delta_i}\, ,
\eea
which is exactly Eq.\ (\ref{eq:I-pN-mG}) with $p_{N+1}=3$.   The third line of Eq.\ (\ref{eq:I-3-3-terms}) is
found by closing the $y\equiv y^-_{N+1}$ contour in the lower half-plane, recalling that all denominators are understood to include $-i\ep$.

\subsection{Generalization for $m_G$ to $n$ identity}

We can generalize the method for $p_{N+1}=3$  to arbitrary $p_{N+1}=n$ outgoing lines. 
We first write the ``$m_G$ to $n$" identity for $I_{n,m_G}$, and then, as in Eq.\ (\ref{eq:I-3-iterative}), identify factors of $I_{n-1,m_G}$
 in each of its terms,
 \bea
I_{n,m_G}\ &\equiv& \sum_{m_G\ge l_{n-1}\ge l_{n-2}\ge\dots\ge l_1\ge 1}\ \prod_{i_n=l_{n-1}}^{m_G}\ \frac{1}{\xi_n-\delta_{i_n}}\ \prod_{i_{n-1}=l_{n-2}}^{l_{n-1}}\ \frac{1}{\xi_{n-1}-\delta_{i_{n-1}}}\times\cdots\times \prod_{i_1=1}^{l_1}\ \frac{1}{\xi_1-\delta_{i_1}} \nn \\
 & =& \sum_{l_{n-1}= 1}^{m_G}\ \prod_{i_n=l_{n-1}}^{m_G}\ \frac{1}{\xi_n-\delta_{i_n}}\ \sum_{l_{n-2}=1}^{l_{n-1}}\ \prod_{i_{n-1}=l_{n-2}}^{l_{n-1}}\ \frac{1}{\xi_{n-1}-\delta_{i_{n-1}}}\times\cdots\times\sum^{l_2}_{l_1=1}\ \prod^{l_2}_{i_2=l_1} \frac{1}{\xi_2-\delta_{i_2}}  \prod_{i_1=1}^{l_1}\ \frac{1}{\xi_1-\delta_{i_1}} \nn \\
  & = & \sum_{l_{n-1}= 1}^{m_G}\ \prod_{i_n=l_{n-1}}^{m_G}\ \frac{1}{\xi_n-\delta_{i_n}}\ \times\ I_{n-1,l_{n-1}} \ .
  \label{eq:mnid}
 \eea
 We assume that this result holds for all integrals $I_{n-1,m_{G'}}$ defined by Eq.\ (\ref{eq:I-pN-mG}) for arbitrary $m_{G'}$, and seek to prove that $I_{n,m_G}$ defined as in this expression is equal to the corresponding integral with $p_{N+1}=n$.
 
Proceeding as in Eq.\ (\ref{eq:I-3-iterative}), our inductive assumption for  $I_{n-1,l_{n-1}}$ implies that it can be replaced in (\ref{eq:mnid}) by the form that results directly from closing the $y$ contour of Eq.\ (\ref{eq:I-pN-mG}) in the lower half-plane,
 \bea 
 I_{n-1,l_{n-1}} \ & = & \sum^{n-1}_{i=1} \prod^{n-1}_{j\neq i} \frac{1}{\xi_j-\xi_i}\ \prod^{l_{n-1}}_{k=1} \frac{1}{\xi_i-\delta_k} \ . 
 \label{eq:m-n-1-res}
 \eea
 We want to show that $I_{n,m_G}$ has the same form,
 \bea 
 I_{n,m_G} \ & = & \sum^{n}_{i=1} \prod^{n}_{j\neq i} \frac{1}{\xi_j-\xi_i}\ \prod^{m_G}_{k=1} \frac{1}{\xi_i-\delta_k} 
 \nn\\
 & = & \frac{1}{2\pi i}\ \int_{-\infty}^\infty d y \ \prod_{l=1}^{n} \frac{1}{\xi_l\ -\ y\ -\ i\ep}\ \prod_{i=1}^{m_G} \frac{1}{y\ -\ \delta_i\ -\ i\ep}\,  . 
 \label{eq:m-n-res}
 \eea

 Inserting Eq.~(\ref{eq:m-n-1-res})  into the third equality of (\ref{eq:mnid}), we get
  \bea 
  I_{n,m_G} &=& \sum_{l_{n-1}= 1}^{m_G}\ \prod_{i_n=l_{n-1}}^{m_G}\ \frac{1}{\xi_n-\delta_{i_n}}\ \left( \sum^{n-1}_{i=1}\prod^{n-1}_{j\neq i} \frac{1}{\xi_j-\xi_i}\ \prod^{l_{n-1}}_{k=1} \frac{1}{\xi_i-\delta_k}\right) \nn \\
   &=& \sum^{n-1}_{i=1}\prod^{n-1}_{j\neq i} \frac{1}{\xi_j-\xi_i}\ \sum_{l_{n-1}= 1}^{m_G}\ \prod_{i_n=l_{n-1}}^{m_G}\ \frac{1}{\xi_n-\delta_{i_n}}\ \prod^{l_{n-1}}_{k=1} \frac{1}{\xi_i-\delta_k} \nn \\
   &=& \sum^{n-1}_{i=1}\prod^{n-1}_{j\neq i} \frac{1}{\xi_j-\xi_i}\left( \frac{1}{\xi_i-\xi_n}\left[\prod_{i_n=1}^{m_G}\ \frac{1}{\xi_n-\delta_{i_n}} - \prod^{m_G}_{k=1} \frac{1}{\xi_i-\delta_k}\right]\right) \ , 
   \label{eq:n-case}
   \eea
   where in the second equality, we have changed the order of the two summations and factored out the products of $1/(\xi_j-\xi_i)$.  The sum over index $l_{n-1}$ is now of the form of Eq.\ (\ref{eq:I-2-result}) with $\xi_n$ and $\xi_i$ replacing $\xi_2$ and $\xi_1$, respectively.   We may therefore apply Eq.\ (\ref{eq:m-to-2-yint}) to get to the third equality.
  
 Taking into account the relative sign,  the denominator  $(\xi_n-\xi_i)$ extends the product over $j$ up to $n$ in the second term of the third equality of (\ref{eq:n-case}). 
 For the first term in the third equality, we only need to use the identity, which can be obtained by closing the contour either in the lower or upper half plane, respectively, for the following integral,
   \bea
   \frac{1}{2\pi i}\int dz \,\frac{1}{z-\xi_n-i\ep}\,\prod_{i=1}^{n-1}\frac{1}{\xi_i-z-i\ep} &= &     \sum^{n-1}_{i=1}\frac{1}{\xi_i-\xi_n} \prod^{n-1}_{j\neq i} \frac{1}{\xi_j-\xi_i} 
   \nn \\
   & = &  \prod_{i=1}^{n-1} \frac{1}{\xi_i-\xi_n} 
 \, .
    \eea
    In this form, we can add the first term to the other $n-1$ terms so that the summation index $i$ in (\ref{eq:n-case}) runs up to $n$ now. As a result, we obtain the complete result for $I_{n,m_G}$ given in Eq.~(\ref{eq:m-n-res}) starting from Eq.~(\ref{eq:mnid}). 
   
\end{appendix}

\end{document}